\documentclass[]{aastex63}
\usepackage{graphicx}
\usepackage{amsmath}
\usepackage{booktabs}
\usepackage{multirow}
\usepackage{float}
\usepackage{lineno}
\usepackage{subfigure}
\usepackage{subfloat}
\usepackage{overpic}
\usepackage{ulem}
\usepackage{longtable}
\usepackage[flushleft]{threeparttable}
\usepackage{hyperref}
\usepackage[hyphenbreaks]{breakurl}
\usepackage{CJKutf8}

\shorttitle{Distinguishing gamma-ray bursts by deep learning methods}
\shortauthors{Zhang et al.}

\begin{document}
\begin{CJK*}{UTF8}{gbsn}

\title{Application of Deep Learning Methods for Distinguishing 
Gamma-Ray Bursts from Fermi/GBM TTE Data}


\def\Tongji{College of Electronic and Information Engineering, Tongji University, Shanghai 201804, China}

\def\HighEnergy{Key Laboratory of Particle Astrophysics, Chinese Academy of Sciences, Beijing 100049, China, \url{libing@ihep.ac.cn, rzgui@tongji.edu.cn}}

\def\Guokeda{University of Chinese Academy of Sciences, Beijing 100049, Beijing, China}

\def\Nanjing{School of Astronomy and Space Science, Nanjing University, Nanjing 210023, China}

\def\Guangxi{Guangxi Key Laboratory for Relativistic Astrophysics, Nanning 530004, China}

\def\XinanJiaoTong{School of Computing and Artificial Intelligence, Southwest
Jiaotong University, Chengdu 611756, China}

\def\HeBeiNormal{College of Physics and Hebei Key Laboratory of Photophysics Research and Application, Hebei Normal University, Shijiazhuang, Hebei 050024, China }

\def\Dezhou{School of Computer and Information, Dezhou University, Dezhou 253023, China}

\def\BeijingNormal{Department of Astronomy, Beijing Normal University, Beijing 100875, China}

\author[0000-0002-8097-3616]{Peng Zhang (张鹏)}
\affiliation{\Tongji}
\affiliation{\HighEnergy{}}

\author[0000-0002-0238-834X]{Bing Li (李兵)}
\affiliation{\HighEnergy{}}
\affiliation{\Nanjing{}}
\affiliation{\Guangxi{}}

\author{Renzhou Gui (桂任舟)}
\affiliation{\Tongji}

\author{Shaolin Xiong (熊少林)}
\affiliation{\HighEnergy{}}

\author[0000-0002-6189-8307]{Ze-Cheng Zou (邹泽城)}
\affiliation{\Nanjing{}}

\author{Xianggao Wang (王祥高)}
\affiliation{\Guangxi{}}

\author{Xiaobo Li (李小波)}
\affiliation{\HighEnergy{}}

\author{Ce Cai (蔡策)}
\affiliation{\HeBeiNormal{}}

\author{Yi Zhao (赵一)}
\affiliation{\Dezhou{}}

\author[0000-0001-5348-7033]{Yanqiu Zhang (张艳秋)}
\affiliation{\HighEnergy{}}

\author{Wangchen Xue (薛王陈)}
\affiliation{\HighEnergy{}}

\author{Chao Zheng (郑超)}
\affiliation{\HighEnergy{}}

\author{Hongyu Zhao (赵宏宇)}
\affiliation{\XinanJiaoTong{}}

\begin{abstract}
To investigate GRBs in depth, it is crucial to develop an effective method for identifying GRBs accurately. 
Current criteria, e.g., onboard blind search, ground blind search, and target search, are limited by 
manually set thresholds and perhaps miss GRBs, especially for sub-threshold events. 
We proposed a novel approach that utilizes convolutional neural networks (CNNs) to distinguish 
GRBs and non-GRBs directly. We structured three CNN models, \textit{plain}-CNN, ResNet, and ResNet-CBAM, 
and endeavored to exercise fusing strategy models. Count maps of NaI detectors onboard Fermi/GBM 
were employed as the input samples of datasets and models were implemented to evaluate their performance 
on different time scale data. The ResNet-CBAM model trained on 64\,ms dataset achieves high accuracy overall, 
which includes residual and attention mechanism modules. 
The visualization methods of Grad-CAM and t-SNE explicitly displayed that the optimal model 
focuses on the key features of GRBs precisely. 
The model was applied to analyze one-year data, accurately identifying approximately 98\% of GRBs listed in 
the Fermi burst catalog, 8 out of 9 sub-threshold GRBs, and 5 GRBs triggered by other satellites, 
which demonstrated the deep learning methods could effectively distinguish GRBs from observational data. 
Besides, thousands of unknown candidates were retrieved and compared with the bursts of SGR J1935+2154 
for instance, which exemplified the potential scientific value of these candidates indeed. 
Detailed studies on integrating our model into real-time analysis pipelines thus may improve 
their accuracy of inspection, and provide valuable guidance for rapid follow-up observations of multi-band telescopes.
\end{abstract}


\section{Introduction}
Gamma-ray bursts (GRBs) are the brightest explosions from distant
galaxies, releasing isotropic energies up to $ 10^{54} $\,ergs in
gamma-rays of prompt emission (e.g. \citealt{Atteia2017ApJ,Minaev2020MNRAS}),
typically peaking in 10–$10^{4}$\,keV energy 
band \citep{Gruber2014ApJS,Preece2016ApJ,Ohmori2019PASJ}, 
and mostly exhibiting a non-thermal spectrum \citep{Band1993ApJ,WangFeifei2020ApJ},
showing intricately irregular pulses from less than a second 
to tens of thousands seconds.
GRBs are generally separated into two classes, long duration GRBs and short
duration GRBs \citep{T90}. 
A large number of observations have supported the theoretical 
predictions that long duration GRBs are produced by collapsing massive stars 
and at least some short duration GRBs originate from binary neutron star mergers (for reviews 
see \citealt{KumarZhang2015PhR,Meszaros2019MmSAI}). 
Still, there are many unsolved mysteries about the origin of GRBs. 
\par
The rapid identification of the GRBs by the space telescope is conducive to guide other 
telescopes for joint or follow-up observations. 
Multi-band follow-up observations are crucial to researching the afterglow, 
host galaxies, and other information of GRBs. 
However, the light curve morphology of GRBs is extremely complex, 
especially the main episode known as the prompt emission phase. 
The light curve is irregular and multi-peak, which makes it difficult  
to identify GRBs quickly and accurately. 
The onboard automatic trigger search algorithm searches triggers 
through the signal-to-noise ratio (SNR) 
of the peak in the light curve, identical to the ground blind 
search algorithm \citep{fermi_blind_search}. 
The coherent search adopted by the target search algorithm requires 
the aid of additional trigger time \citep{fermi_target_search}.
\citet{relate_search_GRB_DTW} designed a method of searching for GRBs from the 
morphology of the light curve. This method used a hierarchical cluster algorithm 
to find candidates, and then compared the light curve of these candidates 
with the typical GRBs in the template bank. 
However, the above methods highly depend on the SNR of the event 
pulses on the detected light curve, which requires manual participation. 
It would miss some weak and inconspicuous events, such as 
sub-threshold bursts\footnote{Note that
sub-threshold GRBs are critical to GW/GRB association, e.g., GBM-190816 
is suggested to be possibly associated with a sub-threshold GW 
event \citep{sub_grb_for_GW, sub_grb_for_GW2}.}. 
\par
In recent years, machine learning (ML) has begun to play a powerful role in various fields. 
ML provides a useful complement to the common paradigm of model-driven data analysis 
through a data-driven approach. The intersection of ML and astronomy is a fairly new paradigm 
that trains artificial intelligence systems on huge amounts of archival data 
to produce high quality results for extracting domain knowledge and facilitating 
new discoveries (for review see \citealt{ML_in_astronomy_overview_2019,ML_in_astronomy_overview_2022,ML_in_astronomy_overview_2022_2}). 
As a sub-field of machine learning, deep learning (DL) excels in discovering effective 
representations of internal patterns and correlations from massive amounts of data automatically. 
DL methods have made great success in computer vision and natural language processing, and 
there are many applications in astronomy and astrophysics 
\citep{deep_wide_faster_fast_transients,DL_Cherenkov_Telescopes,DL_astro_white_paper}. 
Deep learning uses computational model consist of multiple processing layers to learn data 
representations at the abstraction level \citep{DL_review}. 
Convolutional neural network (CNN) is a commonly used deep learning algorithm that 
automatically extracts features from high-dimensional data. The CNN is useful for image 
classification because it learns the underlying features of samples layer by layer 
and retains the spatial relations between pixels. 
Recently, deep learning has been developing rapidly and gained widespread attention and application 
in the field of astronomy and astrophysics, and shown a growing trend in hot topics of astronomy research. 
A large number of research efforts have been made for digging knowledge depth from huge amounts of 
observation data, e.g., gravitational wave detection \citep{DL_detect_classify_GW}, gravitational lensing identification \citep{CNN_gw_lens}, exoplanet search \citep{ml_search_exoplanets}, transient source classification \citep{relate_FRB_deep_learning2}, solar flare prediction \citep{dl_forecast_solar_flare}, background estimation \citep{crupi_GRB_RNN}, etc. 
The image classification problems puzzled astronomers for a long time, CNN algorithms are a type of neural 
network that bring us closer to understanding the intrinsic characteristics within abundant images. 
\citet{CNN_gw_lens} described a CNN technology to perform fast and automated gravitational lens analyses. 
Moreover, relying on the quality of the dataset, recent studies have shown that CNN methods could be used for 
inferring galaxy cluster masses directly from mock images and galaxy member dynamics rather than on any 
other assumptions. \citet{DL_infer_galaxy_cluster_masses} applied the CNNs that trained by using simulated 
intensity maps to predict the galaxy cluster masses from the real Planck maps, and obtained an results overall in agreement with masses that estimated by Planck and determine. 
The signal search of bursts is another puzzling problems in time domain astronomy, and the CNNs are used 
to analyze data in several contexts. 
\citet{Object_detect_celestial} used the object detection method to detect celestial objects candidates of 
from the image with photon flux distribution, which achieves over 90\% completeness for targets. 
\citet{CNN_search_FRB_81} attempted to apply an algorithm based on ResNet to search for weak Fast 
radio bursts from the data archive of four years observations of Parkes radio telescope. 
Finally, 81 new candidates that were difficult to discover manually were identified. 
\citet{DL_detect_GRB_intensity-map} presented a CNN approach for detecting GRBs by classifying 
the AGILE-GRID intensity maps database, the GRB detection capability was improved dramatically. 
Overall, large amounts and high quality observed or simulated data were used to train and validate CNN models which achieved impressive results. 
\citet{DL_autodencoder_detect_grb} applied the CNN auto-encoder to reconstruct the background-only 
light curves, and the GRBs were searched by reconstruction error exceeding the threshold. 
This method recovered 72 GRBs which are not present in AGILE catalog and was implemented in 
real-time analysis pipeline. 
Astronomy is rich with full raw information within time series data and image-based types 
of data that are well suited to analysis via CNNs in particular. 
Large amounts and high quality observed or simulated data were used to train and validate CNN models 
that impressive results were achieved as due to their extensive employ.

\par
Until January 2022, the Fermi Gamma-ray Burst Monitor (GBM) has detected over 3200 GRBs.
Such a large amount of data allows us to directly use the observed 
real GRBs to train the deep learning model. 
Furthermore, the dimensionality reduction method is utilized to visually analyze 
the distribution differences among the various types of detected GRBs.
The dimensionality reduction method transforms the raw data from a high-dimensional space 
to a low-dimensional space, enabling easier visualization and evaluation.
\citet{tsne_classify_grb1} employed the dimensionality reduction method of 
t-distributed Stochastic Neighbor Embedding (t-SNE) to achieve clear separation between 
short bursts and long bursts from the light curves of the Swift/BAT satellite.
\citet{steinhardt_2023_tsne_classify_grb} utilized the t-SNE method to analyze 
the data from three GRB satellites and discovered uncommon types of GRBs in addition to the long and short bursts.
\par
Inspired by the above research, we design an image classification method 
by CNN to distinguish GRBs. 
Based on observations from Fermi/GBM spanning over 13 years, we construct a training set, 
validation set, and test set, as described in Section \ref{sec:data set}.
Section \ref{sec:methods} presents the structure of three models and their training procedures. 
Additionally, we introduce analysis methods for feature visualization (Grad-CAM) and 
dimensionality reduction visualization (t-SNE).
Section \ref{sec:result} shows the test results of each model and demonstrates 
the application of the optimal model in identifying GRBs from one year of Fermi data.
Section \ref{sec:discussion} discusses the performance and practical application effects of 
the model, and presents the conclusion.

\section{Data Set}
\label{sec:data set}
For training our deep learning models, we need to build the data set 
that consist of GRB category and non-GRB category. 
The NaI detectors of GBM on board the Fermi telescope have  
detected 3083 GRBs to the end of June 2021. 
The Time-Tagged Event (TTE) format data of these GRBs 
has been published online\footnote{\url{https://heasarc.gsfc.nasa.gov/FTP/fermi/data/gbm/burst/}}. 
We download all of the GRBs data, which contains the data of all triggered detectors for each GRB. 
In order to generate the samples of the GRB category, the data needs to be filtered.
In Fermi GRBs area, the distribution of $T_{90}$ is researched \citep{fermi_gbm_fouth_catalog}. 
Over 90\% GRBs are less than 100 seconds. 
The entire burst sample are needed, accordingly the data 
should include the burst phase and 
the background before and after the burst. 
For the Fermi/GBM detected GRBs, its published burst TTE data overlaps the whole 
$T_{90}$ period of each GRB event and covers from roughly 20\,s pre-trigger to 300\,s post-trigger. 
Considering computing efficiency and performance of the deep learning models, 
we take a 120 seconds length for each GRB sample, which contained the complete burst 
period and the background for at least 10 seconds before and after the burst event. 
There are only about 3000 GRBs left, which seems to be a tiny data set for training 
deep learning model. 
Each GRB event was detected by multiple NaI detectors of Fermi/GBM, 
and each event appears slight differences in different detectors due to the direction 
and response of the NaI detectors. 
Consequently the event that triggered by each single detector would be recognized 
as an individual GRB event. 
It is possible to multiply the number of GRB samples. 
We take the signal of each trigger detector as an independent GRB sample. 
Finally, we obtain 6330 GRB samples which might be a sufficient GRB set 
for training and testing.

We mainly use the daily data of Fermi/GBM for building the non-GRB set. 
All the Fermi/GBM daily observations 
are also published\footnote{\url{https://heasarc.gsfc.nasa.gov/FTP/fermi/data/gbm/daily/}} 
in TTE format. 
We should ensure the data do not include the GRBs. 
The daily data products are composed of all GBM detector observations 
continuously in hour, regardless of whether a burst occurred. 
We download a number of daily data products after filtering out the burst 
TTE data of Fermi GRBs. 
We randomly choose the time and ID of NaI detectors to produce non-GRB data. 
A total of 4000 daily TTE files that do contain the time period of GRBs were downloaded.
Sequentially, 10 segments of non-overlapping 120 seconds long data are extracted 
from each choosing data, and 40000 non-GRB events are obtained in total. 
All these non-GRB events are divided into three sets according to time. 
It is important to maintain the quantitative balance between the 
sample categories. 
Each of the three sets is randomly arranged, and we randomly select
10000 non-GRB events as non-GRB samples.

\begin{figure*}[htbp]
\hspace{-1cm}
\begin{overpic}[width=10cm]{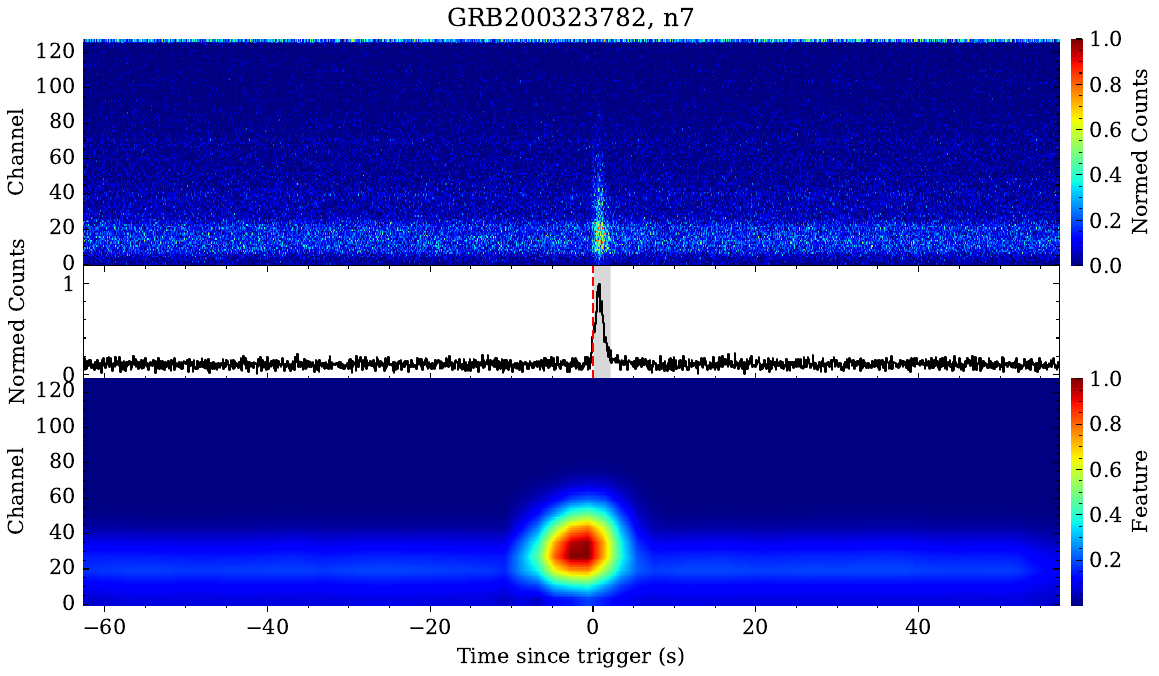}
  \put(-2,55){}
\end{overpic}
\begin{overpic}[width=10cm]{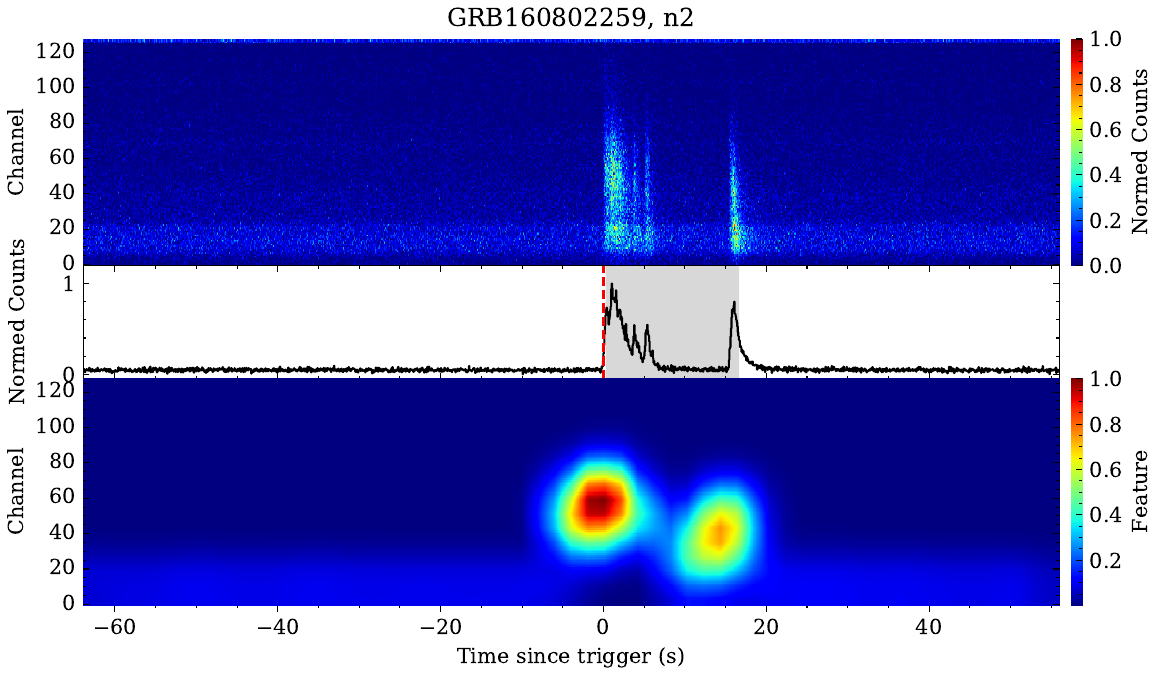}
  \put(-2,55){}
\end{overpic}

\hspace{-1cm}
\begin{overpic}[width=10cm]{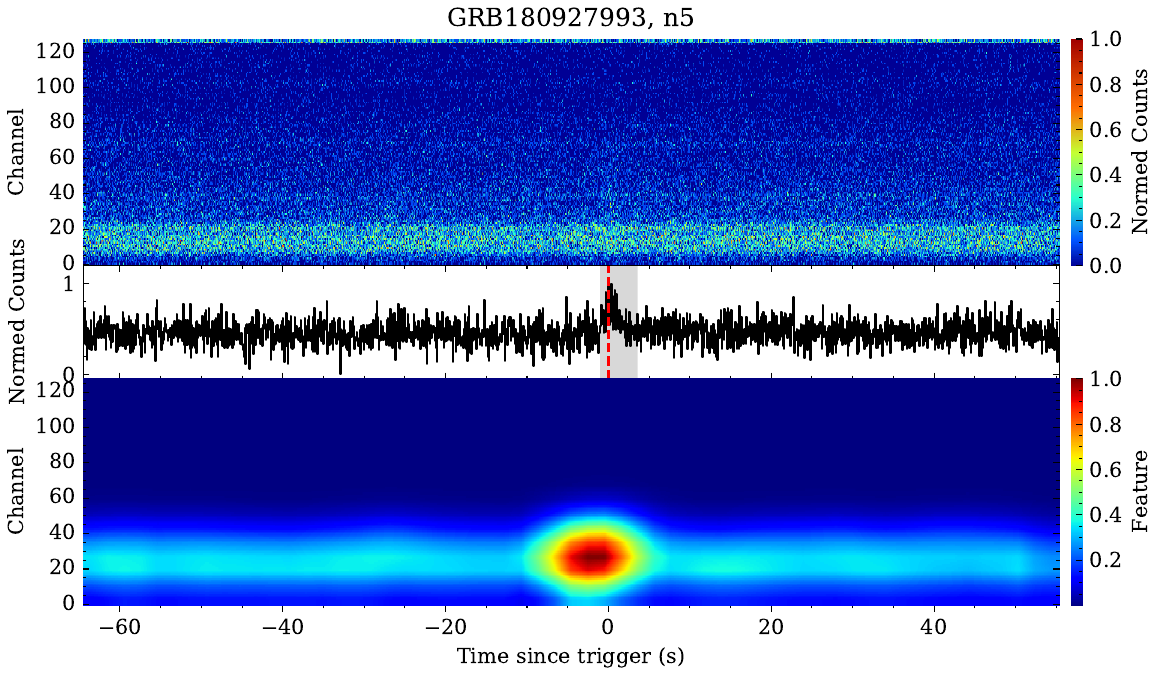}
  \put(-2,55){}
\end{overpic}
\begin{overpic}[width=10cm]{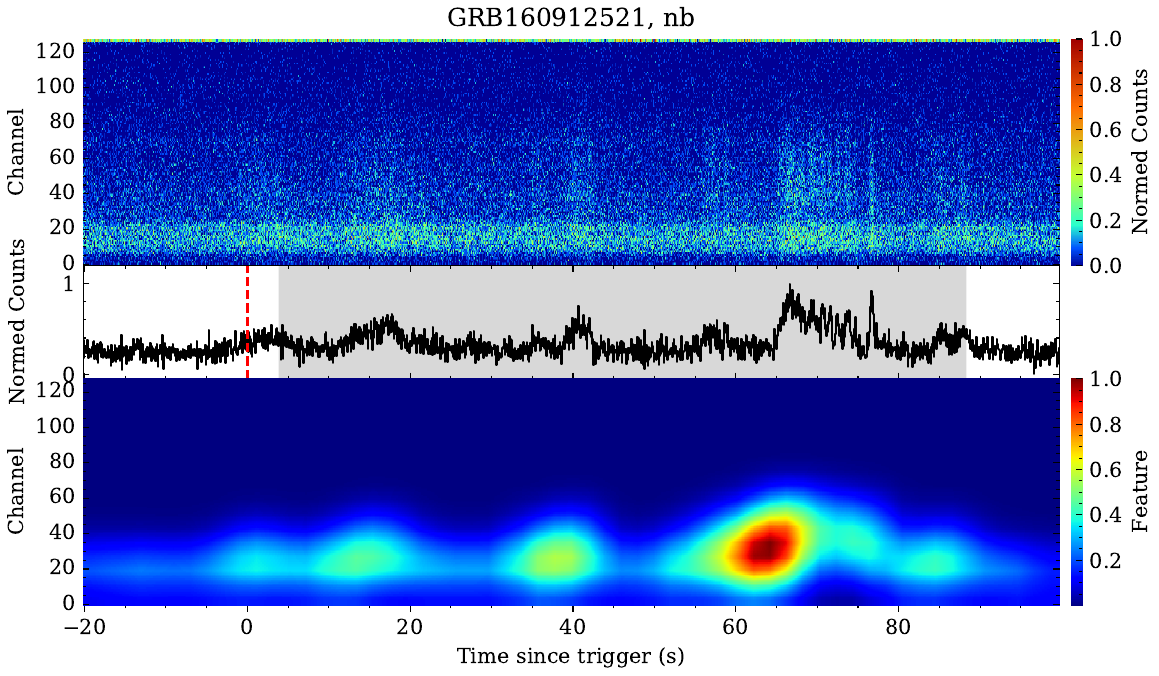}
  \put(-2,55){}
\end{overpic}
\caption{Four representative samples from the data set. 
The GRBs with single peak (GRB200323782, n7), two peak (GRB160802259, n2), low SNR 
(GRB180927993, n5), and complex structure (GRB160912521, nb) are represented. 
For each sub-ﬁgure, the top panel shows the normalized count map (model input), the central panel 
shows the normalized light curve of full energy band, while the bottom panel 
is the feature heat-map generated by Grad-CAM method. The red dashed line represents 
the trigger time of Fermi and the gray area indicates the Fermi-$T_{90}$.}
\label{fig:lc_example}
\end{figure*}

The count maps of GRBs contain essential information about their physical 
features and emission mechanisms, see Figure \ref{fig:lc_example}. 
Both GRB samples and non-GRB samples are composed of 
total 128 channel data covering energies from 8\,keV to 2\,MeV. 
Notice that we do not use the detector response function to convert 
channel to energy. 
The count rate data of the whole samples are not corrected by dead time.
There exist a significant variety of burst signals between different 
time scales \citep{relate_diff_bin_explain}. 
Generally, multiple time scales of 16, 32, 64, 128, and 256\,ms are implemented 
in searching trigger \citep{fermi_gbm_fouth_catalog,fermi_target_search}. 
In our data pre-processing, the time scales of 256, 128, and 64\,ms are respectively adopted. 
To eliminate the magnitude difference between samples, 
each count map is individually normalized. 
It's worth noting that we used a method known as min-max normalization in this 
procedure to normalize the data values of counts in each count-map. 
Each count value was converted to a value between 0 and 1 referring to a 
formula: New value = (value – min) / (max – min) * 1.
The three data sets of training set, validation set, and test set are essential in 
deep learning method. 
The training set is used to fit the models, and after each training epoch, 
the validation set is applied to verify the performance of the current model. 
The test set is adopted to assess generalization of the final optimal model. 
To avoid data confusion, we divide the samples into the above data sets 
based on time period, see Table \ref{table:data set} for detail. 
The ratio of the sample numbers in these three data sets is reasonable.

\section{Methods}
\label{sec:methods}

\subsection{Architecture of Neural Networks}
\label{sec:nework_architectures}
\begin{figure*}[htbp]
\centering
\includegraphics[width=15cm]{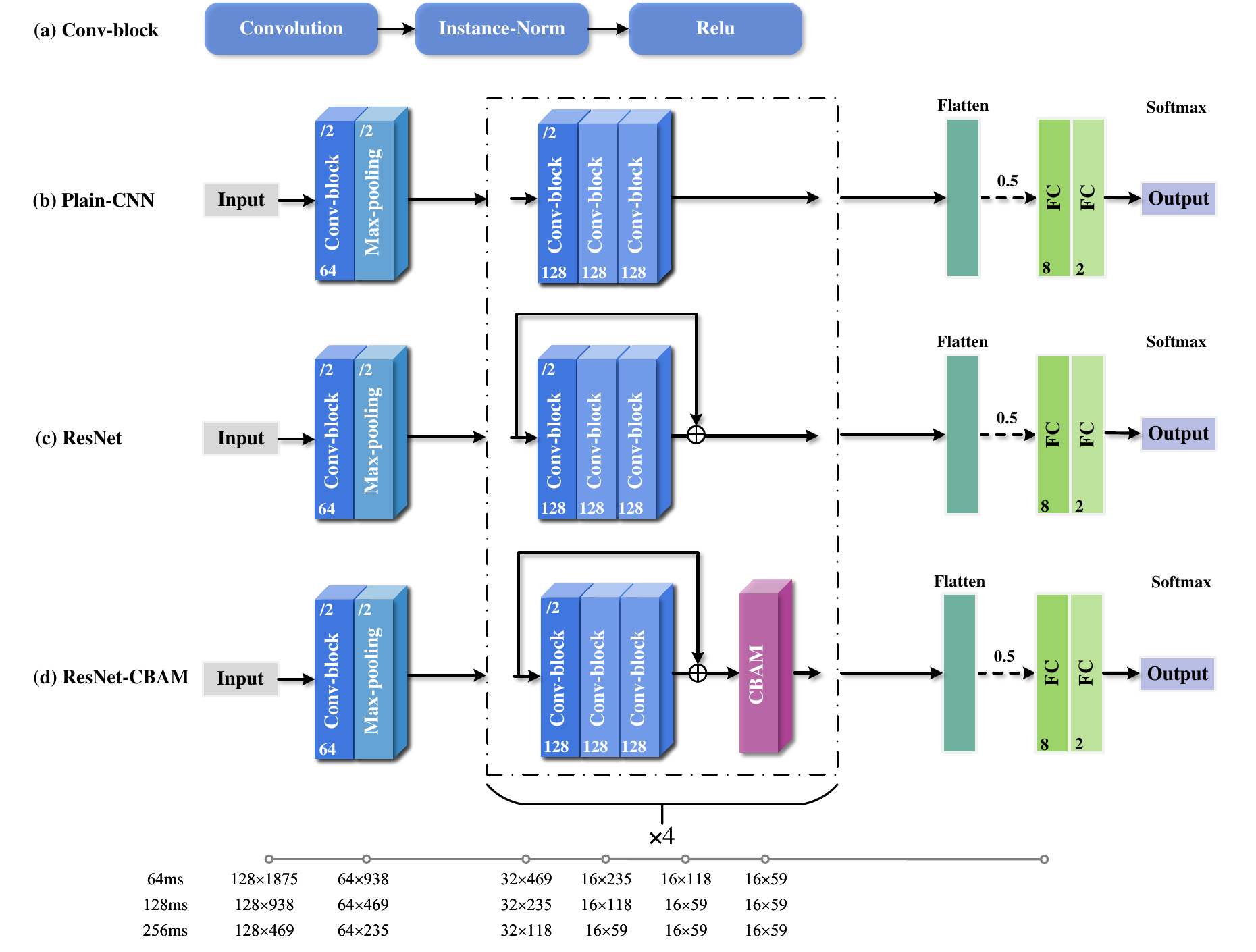} 
\caption{
Schematic diagram of our CNN architectures. 
The numbers at the top and bottom of Conv-block denote the convolutional stride (default is 1)
and number of convolutional kernels, respectively. 
The number at the bottom of the figure describes the variation of the length
and width of the feature maps.}
\label{fig:network_architectures}
\end{figure*}

The convolution block (Conv-block) is the base block 
for all the models in our work, comprising a convolutional layer, 
an instance normalization layer, an activation function, 
see Figure \ref{fig:network_architectures}(a). 
We use multiple convolutional kernels of size $3\times3$ and of default 
stride 1 in the convolutional layer to extract features from input samples.  
The Instance Normalization layer, proposed by Ulyanov \citep{Instance_norm}, 
was utilized to normalize the feature maps, thereby constraining the convergence of the model. 
Nonlinear factors are incorporated into the model through the activation function 
to enhance its expressiveness.
We adopt an effective activation function called Rectified Liner Unit (Relu) 
presented by \citet{relu}. 
The advantages of the \textit{Relu} function include addressing the vanishing gradient problem, faster computation speed, and sparse activation. 
This activation function is widely used in deep learning model \citep{model_resnet, model_block_CBAM}.

\par
Based on the Conv-block, we construct a general CNN model named as \textit{plain}-CNN, 
as shown in Figure \ref{fig:network_architectures}(b). 
The initial component of plain-CNN comprises a Conv-block with 128 convolutional 
kernels and a max-pooling layer, both with a stride of 2.
The primary objective of setting the stride as 2 is to reduce the size of the 
feature map, thereby simplifying the model's complexity and enhancing computational efficiency.
In the subsequent section, with the aim of extracting deeper features, we sequentially 
stack 4 convolutional units, each comprising three Conv-blocks, 
with every Conv-block containing 128 convolutional kernels. 
The final component of the \textit{plain}-CNN model consists of a flatten layer, a dropout layer, and a classifier.
The flatten layer sequentially flattens the features output from the fourth 
convolutional unit, thereby converting the multidimensional input into a one-dimensional form.
We add a dropout layer which is proposed by \citet{dropout}. 
During model training, this layer could randomly deactivate neurons with a 
probability of $p$ to alleviate model overfitting, and we conservatively set $p$ to 0.5.
As the nonlinear classifier, it consists of two fully connected layers (FC) with 8 and 2 neurons, respectively. 
The classification results of the input samples are presented as probabilities. 
We employ the Softmax function to calculate the probabilities for binary classification, whose formula is
\par
\begin{equation}
Softmax(z_{i})=\frac{ e^{z_{i}} }{ \sum_{c=1}^{C} e^{z_{c}}, }
\end{equation}
\par
where $z_{i}$ is the output of the classifier. $i$ denotes 
the category index and the total number of categories is $C$.
\par
Deep neural networks that stack a large number of convolutional layers may suffer from 
the degradation problem, which leads to a decrease in model performance. 
\cite{model_resnet} proposed a residual module to mitigate the degradation problem of network. 
The residual module establishes a shortcut (skip) connection between a shallow layer 
and a deep layer to achieve identical mapping, which enables the model to ignore redundant layers.
The residual module also enables fast convergence in the early stages of model training,
and facilitates the training of deeper networks more effectively.
For solving the problem of the vanishing/exploding gradient, the residual blocks were 
introduced into the architecture of plain-CNN model making an extension and enhancement network model, 
known as the residual network (ResNet).
Residual blocks were stacked together to make a ResNet which model is characterized by 
adding shortcut connections (+) to each convolutional unit of the second part of 
\textit{plain}-CNN model, whereas other structures remain consistent with it, 
see Figure \ref{fig:network_architectures}(c). 

\par
One key focus of CNN architecture design is the attention mechanism, which achieves 
feature selection by assigning weights to features based on their significance. 
\cite{model_block_CBAM} proposed a lightweight and plug-and-play attention module, 
the Convolutional Block Attention Module (CBAM), which performs attentive operations 
jointly in spatial and channel dimensions and separately refines feature values using spatial 
attention and channel attention of input features. 
The feature maps and attention weights are multiplied to achieve adaptive feature refinement, 
which significantly enhances the feature values.
On the basis of ResNet model, we add a CBAM module to the end of each convolutional unit, 
forming the ResNet-CBAM model, see Figure \ref{fig:network_architectures}(d). 
The CBAM moduls were introduced into the architecture of ResNet model, enabling the model 
to selectively focus on critical features, enhancing the model's representational capacity 
and recognition performance consequently. In contrast, the \textit{plain}-CNN and ResNet could 
emphasize deep feature extraction primarily, while the ResNet-CBAM further enhances the feature representation capacity of model through attention mechanisms. 
The ResNet-CBAM model was indeed structured to serve supplementary and augmenting components to ResNet.

\subsection{Training and Optimization}
\label{sec:train_model}

\subsubsection{For Single Network}
\label{sec:triain single network}
The training of the original model requires the initialization of parameters for each hidden layer at the beginning.
Here we set the initial parameters of all hidden layers as a truncated 
normal distribution which is generated by the \textit{he\_normal} \citep{param_he_normal}. 
To reduce memory consumption, the training sets are divided into multiple batches to train our models. 
During each training epoch, only one batch of samples is input into the model, 
and the classification probabilities are output in turn. 
Both GRB set samples and non-GRB set samples are randomly shuffled and labeled with a sequence number. 
Then, adjacent 32 samples are grouped as a batch based on their sequence number, and any remaining 
samples with fewer than 32 are also used as one batch.
This process is known as forward propagation. 
Loss function plays a crucial role in deep learning model, and it is used to guide model optimization and parameter updates. 
The cross-entropy loss function are commonly used for classification 
models \citep{cross_loss1,cross_loss2}. 
We choose categorical-crossentropy as the loss function to measure the discrepancy 
between the predicted output and the expected output.
This loss function is defined as 
\par
\begin{equation}
Loss= - \sum_{c=1}^{C} y_{c} \cdot \log p_{c}
\end{equation}
\par
where $C$ is total number of categories and $y$ is the real label of the sample. 
$p$ is the predicted value of the model output by the Softmax function. 
\par 
The purpose of iterative training is to find the optimal parameter settings for 
the hidden layers in order to minimize the loss using the gradient descent algorithm. 
This algorithm calculates the gradient of various parameters based on the loss.
The negative gradient, combined with a variable known as the learning rate, is used to update the parameters.
This process is known as backward propagation. 
During each training epoch, every batch of samples from the training dataset goes through
forward and backward propagation sequentially.
We choose Adam that proposed by \citet{adam_optimiser} as the optimizer to accelerate this gradient descent. 
Setting a non-fixed learning rate at each stage of the iterative optimization would facilitate 
finding the minimum loss quickly. At the early stage, a large learning rate was 
set to accelerate parameter optimization, and then we gradually reduce the learning rate 
to search for optimal parameters. 
Many researches set learning rate as 0.001 for deep learning models \citep{DL_detect_GRB_intensity-map,dl_set_lr_3,dl_set_lr_2}.
We conducted experimental tests to assess the impact of learning rates 
of 0.01, 0.001, and 0.0001, respectively. 
We set 0.001 as the initial learning rate for our models. 
If the training loss does not decrease after 20 consecutive epochs, we reduce the learning rate 
by a factor of 2 until it reaches 0.00005. 
The early stopping mechanism is widely adopted prevent the overfitting issue \citep{dl_early_stop0,dl_early_stop2,dl_early_stop1,dl_early_stop3}. 
The training process would be terminated when the validation accuracy does not increase for 40 consecutive epochs. 
The learning curve is utilized to describe the evolving performance of a model on training data over experience or time. 
It assists in analyzing the training effectiveness of model, diagnosing whether the model is experiencing in overfitting issues or underfitting issues, reducing the computational complexity of model training and hyperparameter tuning, and important for the selection of an appropriate model. 
The parameters settings corresponding to highest validate accuracy in the training 
epochs are saved and adopted as our final model. 
Our models are implemented using Keras\footnote{\url{https://keras.io}} with the 
TensorFlow\footnote{\url{https://www.tensorﬂow.org/}} backend. 
We train ResNet-CBAM with batch size of 32 per epoch, which takes around 
130\,s on a computer with one GPU (NVIDIA GTX-1080Ti).

\subsubsection{Network Fusion}
\label{sec:network_fusion}
\begin{figure*}[htbp]
\centering
\includegraphics[height=6cm]{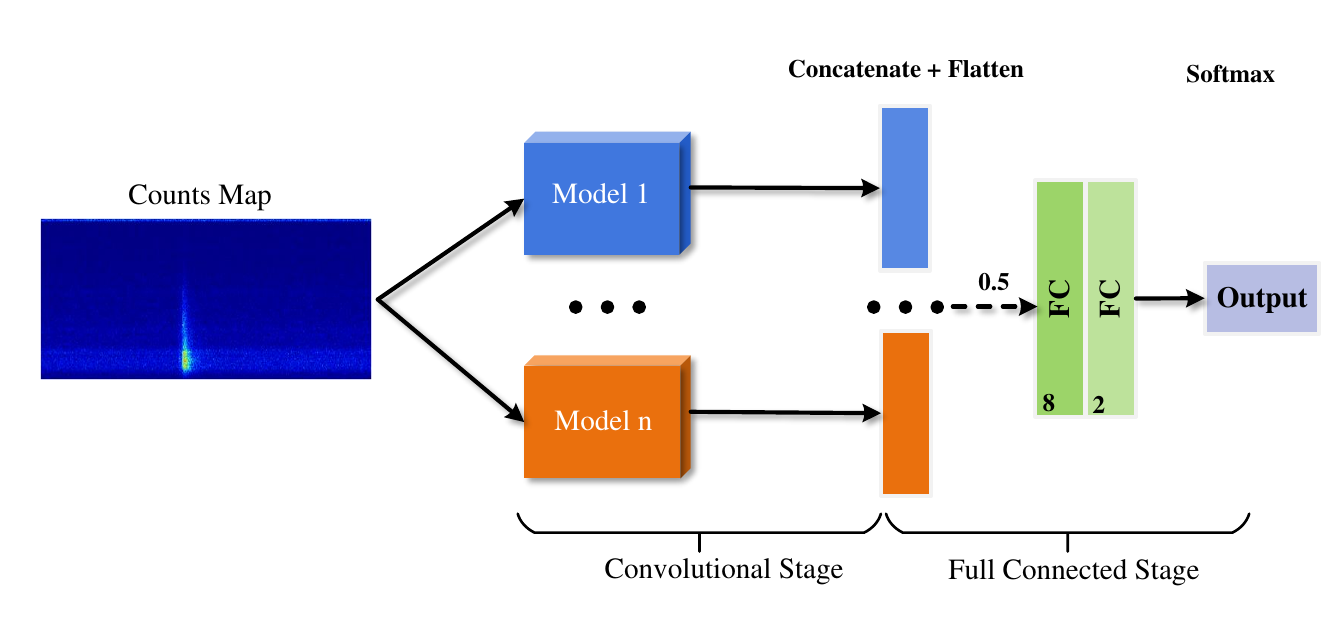} 
\caption{Overall architecture of fused model.} 
\label{fig:network_fusion}
\end{figure*}

The feature extracting capabilities and characteristics of each single network are different. 
Therefore, effective fusion algorithms are also introduced to usually improve accuracy.
The fusion model takes the output from the multiple single network algorithms 
and determines the learning accuracy. 
Our model fusion strategy is shown in Figure \ref{fig:network_fusion}. 
All the fused models include a convolutional stage and a full connected stage. 
At the convolutional stage, the data set is inputted to the respective single network model 
that has already been trained (see \ref{sec:triain single network}), and undergoes a complete 
training with frozen parameters.
Then, the output features of multiple models are concatenated at the fully connected stage 
and flattened. 
Here, we use two new FC layers with 8 and 2 neurons, respectively.
For the fused models, we only train and update the parameters of the two new FC layers, 
whose training procedure is the same as Section \ref{sec:triain single network}. 
The labels of our fused models are listed in Table \ref{table:network_fusion}.

\subsection{Visualization Analysis}
\label{sec:visualization analysis}

\noindent \textbf{Feature Visualization:} Deep learning algorithms have been widely applied in various scenarios, 
but their decision-making processes are opaque to humans. Therefore, these models are often perceived as black boxes.
It is necessary to quantify and visualize the features that extracted by Neural Networks. 
\citet{model_Grad-CAM} proposed an approach to provide visual explanations for deep learning algorithms 
labeled as Gradient-weighted Class Activation Mapping (Grad-CAM). 
Grad-CAM produces a heat-map that highlights the crucial regions of an image by 
using the gradients of the target features from the final convolutional layer. 
This means that the feature maps from the final layer are extracted, and each channel in 
those features is weighted with the gradient of the class with respect to the channel.
Referencing to \citet{model_Grad-CAM}, the heat-map of features, $ L^{c}_{Grad-CAM}$, 
are computed in this work, and presented in the bottom of each panels in 
Figure \ref{fig:lc_example} as examples. 
The most significant feature is the maximum value of the mapping-curves of features  
obtained by summing the heat-map of features along the channel axis. 

\par
\noindent \textbf{Dimensionality Reduction Visualization:} The feature map from the 
final convolutional layer is a high-dimensional and complex dataset, 
which makes it challenging to distinguish and compare bursts, as it is difficult to determine which information is most important.
Reducing high-dimensional data and representing it in 2D or 3D could directly visualize the pattern of data distribution. 
\citet{relate_tsne_method} proposed a non-linear dimensionality reduction algorithm, 
T-distributed Stochastic Neighbor Embedding (t-SNE), which has great application value in big data analyzing. 
Here we apply the t-SNE technique to analyze the feature map of test set, 
and use the \textsf{sklearn.manifold.TSNE} method to implement t-SNE by 
scikit-learn\footnote{\url{https://scikit-learn.org/}}. 
As a comparison, the count map of samples in test set without normalized preprocessing are also analyzed. 
The custom parameter \textit{Perplexity} specifies the importance of local or global structure, 
which generally represent the number of nearest neighbors of each data. 
The custom parameter \textit{Init} represents the initialization of embedding. 
We respectively set the \textit{Perplexity} (ranging in 5, 20, 40, 60 and 80) and \textit{Init} (ranging in $pca$ and $random$) to check the classification stability of t-SNE, 
which refers to the hyper-parameter settings in these publications\citep{tsne_classify_grb1,tsne_classify_grb2,steinhardt_2023_tsne_classify_grb}. 
By visualizing and analyzing the effect of the distinction between the two category data, 
we chose \textit{Perplexity} with 20 and \textit{Init} with 'random' as the optimal setting, while the other hyper-parameters take 
the default values of the \textsf{sklearn.manifold.TSNE} method. 

\section{Result}
\label{sec:result}

\subsection{Model Performance}
\label{sec:model_performance} 
\begin{figure*}[htbp]
\centering
\includegraphics[width=17cm]{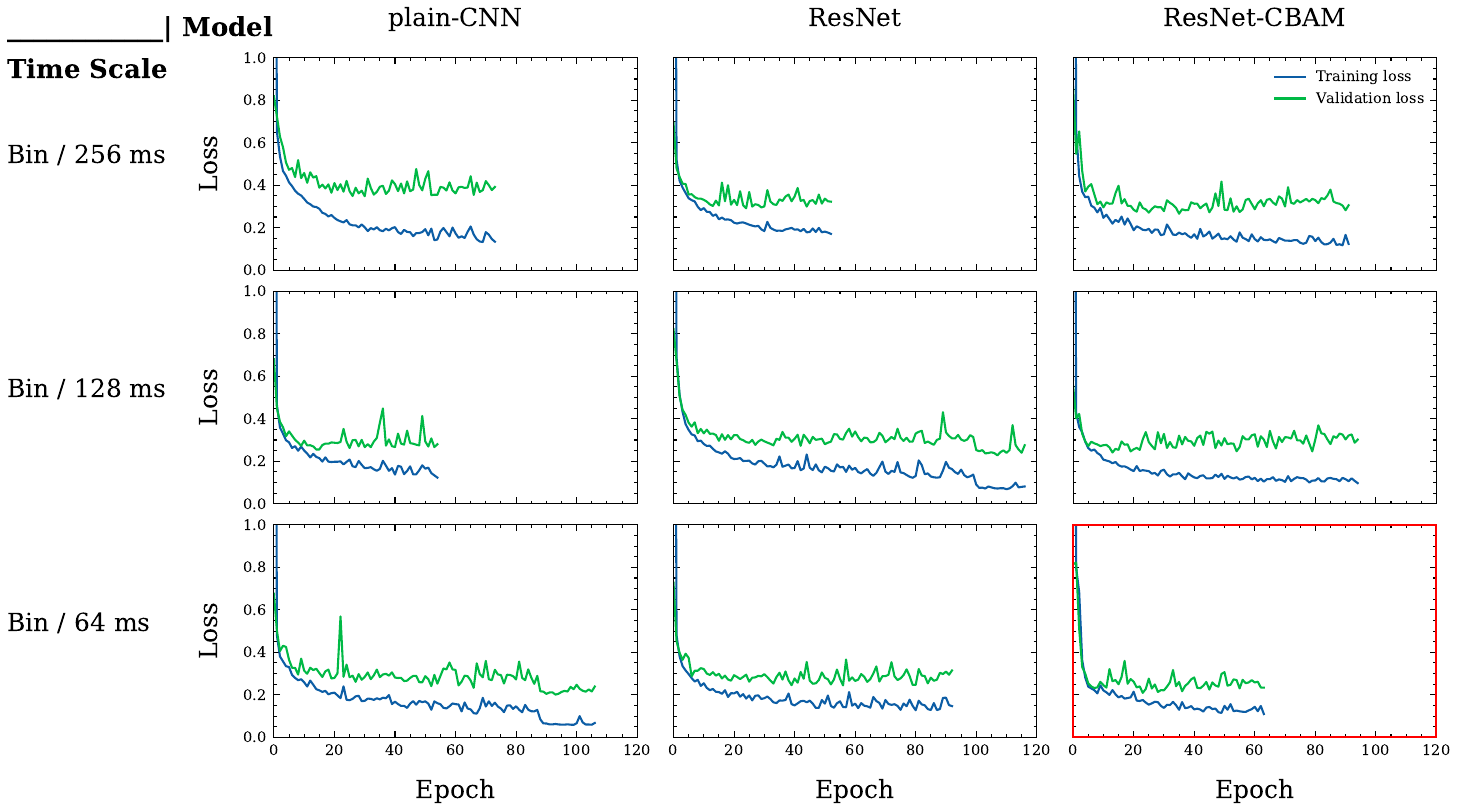} 
\caption{The learning curve of nine single network models. Due to the training mechanism of early stopping, the epoch at which the model stops varies.} 
\label{fig:learning_curve}
\end{figure*}

\begin{figure*}[htbp]
\centering
\includegraphics[height=15cm]{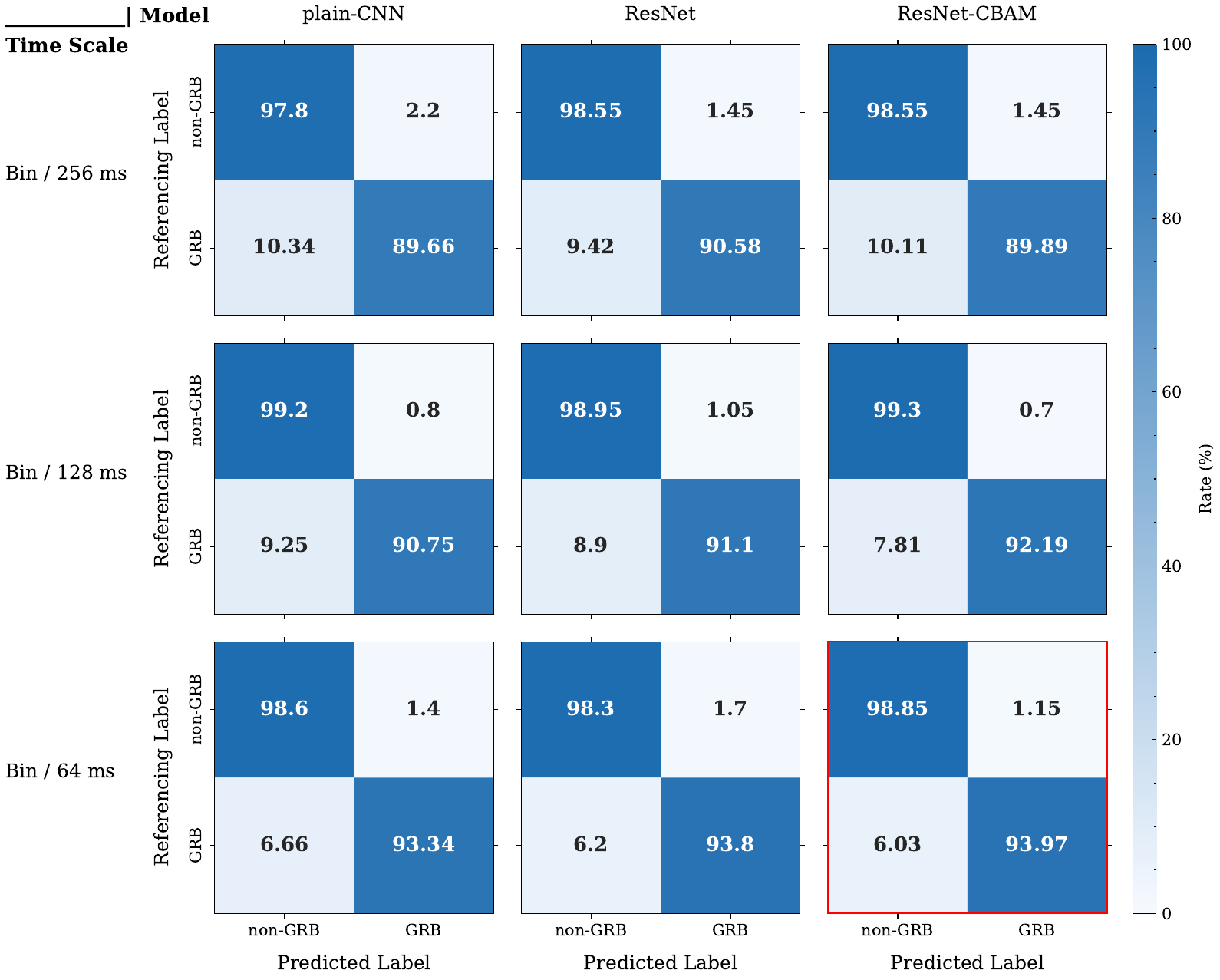} 
\caption{The confusion matrices of nine single network models.} 
\label{fig:confusion_matrix}
\end{figure*}

\begin{figure*}[htbp] 
\centering 
\includegraphics[width=15cm]{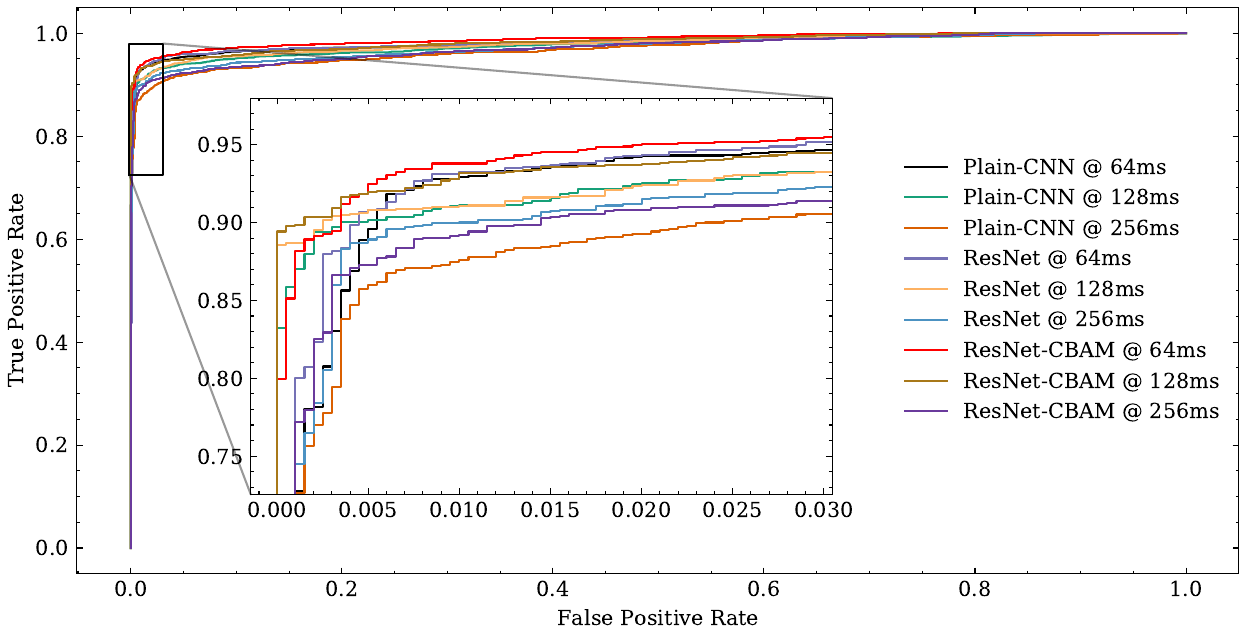} 
\caption{The ROC curves depicts the 
True Positive Rate ($\frac{TP}{TP+FN}$) versus the False Positive Rate ($\frac{FP}{FP+TN}$). 
} 
\label{fig:model_ROC}
\end{figure*}

Three data sets are constructed from over ten years observation data of Fermi/GBM as shown in 
Table \ref{table:data set}, and their time bins are 256, 128, and 64\,ms, respectively. 
We build three single network models, \textit{plain}-CNN, ResNet and ResNet-CBAM, 
and train them on each data sets. 
The learning curves for model training are displayed in 
Figure \ref{fig:learning_curve} which reveals partial overfitting of our models. 
We prevent the overfitting issue by truncating model training procedure through the use of early stopping mechanisms. 
Those models are evaluated using test sets with different time bins, 
and their performances are illustrated by four metrics, \textit{Accuracy}, 
\textit{Precision}, \textit{Recall} and \textit{F1-score} respectively, see Appendix \ref{sec:metrics}. 
The performance metrics of each model in the training set, validation set, and test set are displayed in Table \ref{table:network metrics}.
The specific recognition rate of the models for the samples of each category 
could be represented by the confusion matrix. 
Figure \ref{fig:confusion_matrix} shows the performance 
of each model on the confusion matrix. 
With incorporation of the residual module and attention mechanism, it strengthens 
the generalization ability of our CNN models. 
It can be quite clearly grasped to discriminate what an optimal model the ResNet-CBAM trained on the 64\,ms data set is. 
The Receiver Operating Characteristics (ROC) is another useful way to compare models from an immediate perspective. 
The area under the ROC curve indicates the ability of classification accuracy. 
The ROC curves of our single network models are shown in Figure \ref{fig:model_ROC}. 
The ResNet-CBAM trained on 64\,ms data set obtains the largest area among those ROC curves, 
which also indicates its powerful classification ability. 
\par
The fused models and their performance are shown in Table \ref{table:network_fusion}. 
The overall performance of all fused models exceed 94\%, 
and is better than single network models. 
However, the improvement is relatively limited. 
The \textit{plain}-CNN+ResNet+ResNet-CBAM model trained with the 64\,ms data set performs well 
among those fused models. 
Whereas, during training and testing, we find fused models consume more resources and exhibit low efficiency.
\par
\begin{figure*}[htbp]
\hspace{-1.3cm}
\subfigure{
	\begin{minipage}{.47\paperwidth}
		\includegraphics[width=10cm]{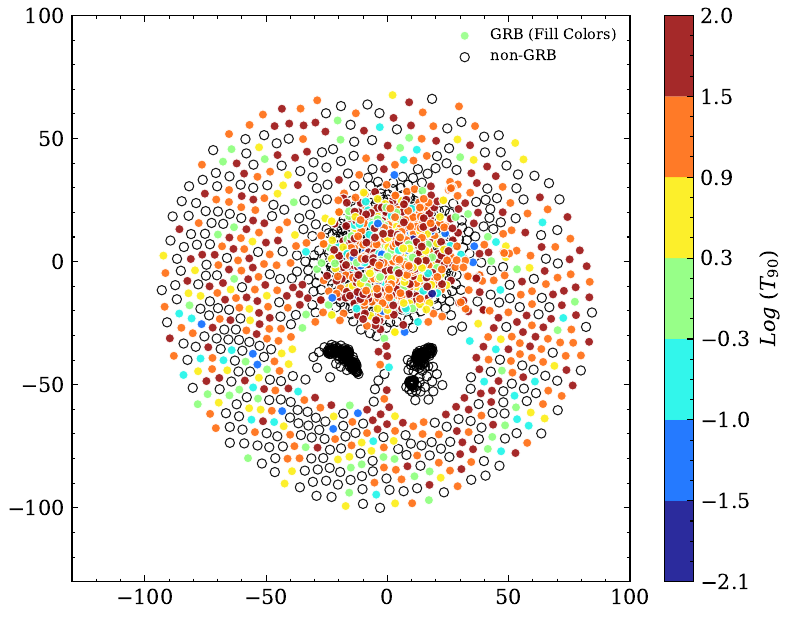}
	\end{minipage}
}
\subfigure{
	\begin{minipage}{.47\paperwidth}
		\includegraphics[width=10cm]{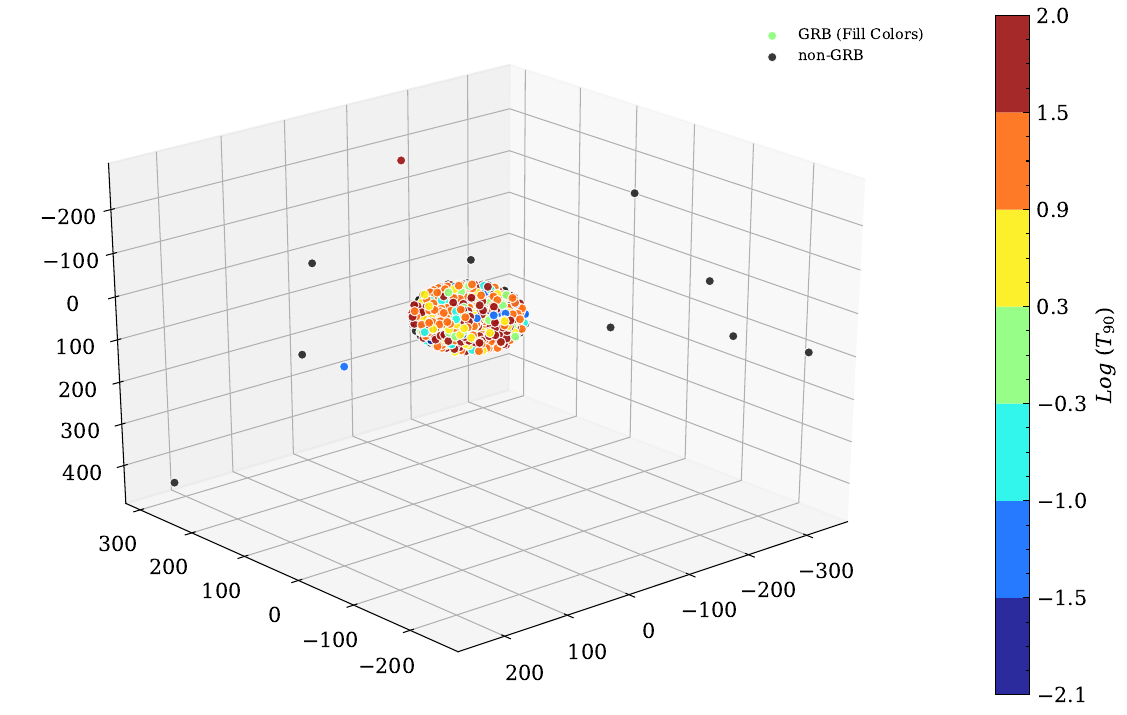}
	\end{minipage}
}
\caption{
The 2D and 3D visualization of the count maps is achieved using the 
dimensionality reduction method t-SNE, with $T_{90}$ information 
embedded for each sample. 
The count maps are the samples in the test set without normalized preprocessing. 
The $T_{90}$ of each GRB sample is derived from the corresponding GRB in 
the Fermi burst catalog. 
It is not able to calculate the duration of the non-GRB samples.
The distributions of primary count maps in feature space for GRB and non-GRB category 
are indistinguishable.
}
\label{fig:sample_testset_tsne_t90_2d_3d}
\end{figure*}
\begin{figure*}[htbp]
\hspace{-1.3cm}
\subfigure{
	\begin{minipage}{.47\paperwidth}
		\includegraphics[width=10cm]{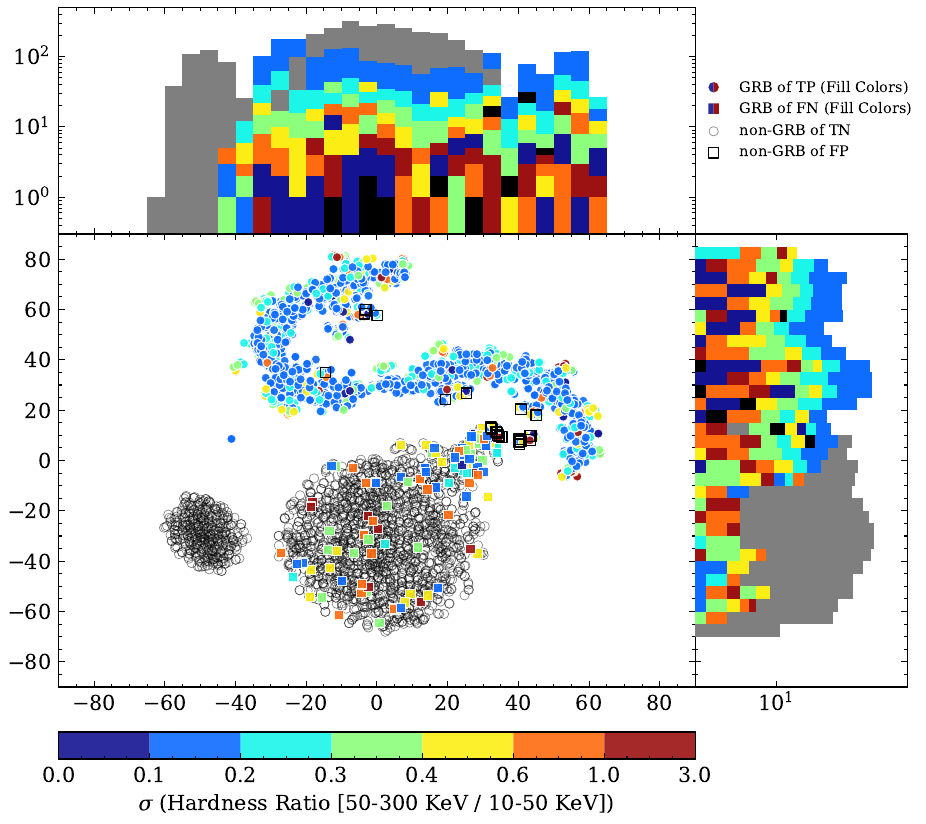}
	\end{minipage}
}
\subfigure{
	\begin{minipage}{.47\paperwidth}
		\includegraphics[width=10cm]{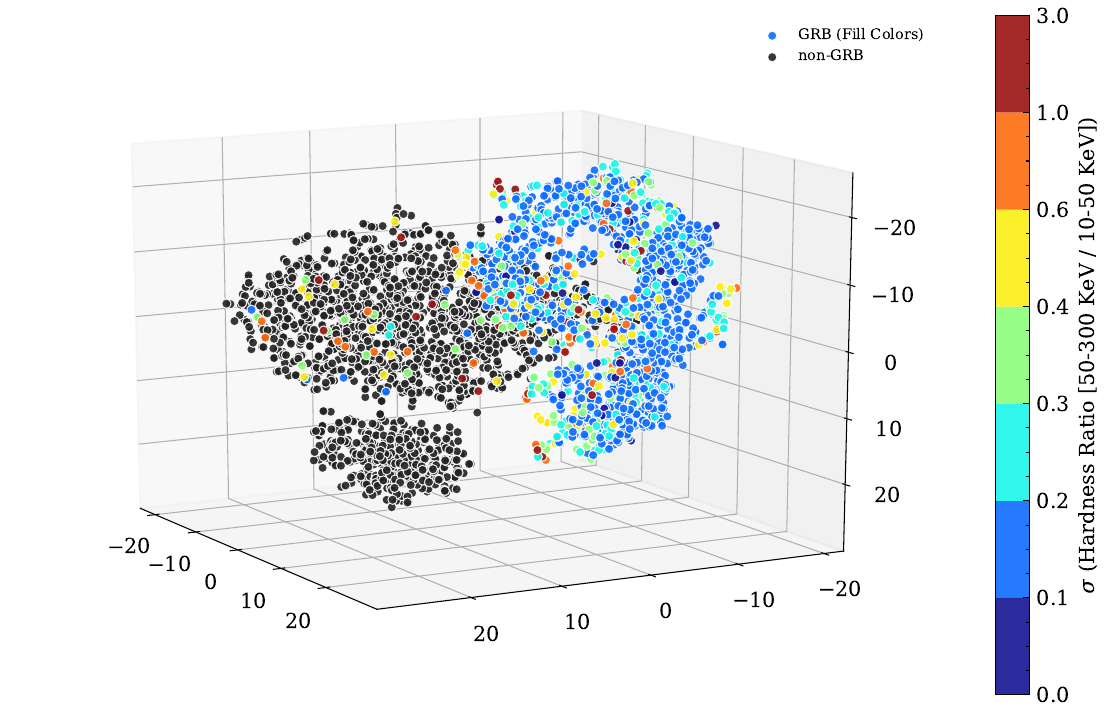}
	\end{minipage}
}
\caption{
The 2D and 3D visualization of the feature maps is achieved using the 
dimensionality reduction method t-SNE, with standard deviation of the hardness 
ratio embedded for each sample. 
The SNR of the GRB of FN samples was relatively low. 
The dispersion of GRB hardness ratios for FN samples is relatively large. 
}
\label{fig:feature_testset_tsne_hardness_std_2d_3d}
\end{figure*}

The Grad-CAM method is selected to perform visualization analysis on features that 
produced by the optimal model. 
This method is effective in locating objects in images based on category conditions. 
The results, namely heat-maps, are shown as the bottom of each panels in Figure \ref{fig:lc_example}. 
Here, we only show four GRBs events corresponding to different morphologies. 
The most interesting areas of the model, namely the most prominent area in heat-map, 
are mainly located at the burst period. 
In addition, the t-SNE technique is applied to reduce the dimensionality of count maps in test set, 
with the 2D and 3D result presented in Figure \ref{fig:sample_testset_tsne_t90_2d_3d}. 
The behavior of the GRB category and the non-GRB category are indistinguishable. 
For GRB category, there are no obvious boundary with various duration. 
Furthermore, feature maps produced by the optimal model, are also applying t-SNE method. 
Figure \ref{fig:feature_testset_tsne_hardness_std_2d_3d} and Figure \ref{fig:feature_testset_tsne_hardness_std_2d_3d} 
depict the results of dimensionality reduction, where the two categories of the test set are distinguished in two separate areas, and there is also no clear distribution of duration among the categories.

\subsection{Application Result}
\label{sec:application} 
We applied the optimal model to search GRBs with a whole year data that observed by NaI detectors 
onboard Fermi/GBM from July 1, 2021 to June 30, 2022. 
A sliding time window was designed to extract data from each daily TTE of NaI detectors continuously, 
with its length of 120 seconds and with a step size of 110 seconds. 
We extract data in reverse when the end of the data file is less than 120 seconds. 
It is important to note here that the time windows for each detector are isochronous for 
the convenience of subsequent joint analysis.
Another important point to note that we selected sliding windows within the range of 
Good Time Interval (GTI) to ensure each time window is completely contained in the GTI. 
The data were processed into count maps, which is consistent with Section \ref{sec:data set}. 
Finally, 3.05 million count maps with bin size of 64\,ms were obtained.
These count maps are inputted into the optimal model, and the classification of GRB or non-GRB is determined.
There are 39,515 classified GRBs that identified as initial events, 
a further screening of those events is needed for searching candidates of GRB, 
and the corresponding heat-maps and mapping-curves of feature were outputted previously. 
\par
The screening process consists of two steps: 1) time filtering: if there are fewer than 
two initial events in the same time window, then these events are excluded. 
The purpose of this step is to ensure that each candidate is detected by at least 
two detectors. 2) location filtering: in each time window, we record the burst moment 
of the most significant feature in mapping-curves of every initial event. 
We sort these moments of bursts and search sequentially for a 20 seconds region from 
the first moment of burst. If there are at least two burst within a 20 seconds interval, 
then they are considered together as a candidate, otherwise repeat this searching process 
with the remaining initial events. 
In other words, there might exist more than one candidate in each time window, 
and it is important to retain as much information about the 
bursts as possible. 
The South Atlantic Anomaly (SAA) is a geomagnetic anomaly region that covers 
the eastern of South America and the south atlantic ocean. 
Fermi/GBM turns off its detectors when crossing over the SAA region to protecti detectors. 
Therefore, data in this period should be excluded. We considered this situation 
and GTI in the data pre-processing process. 
However, many events are still located in the SAA region, so we exclude 45 events. 
After the aforementioned screening process, a total of 1963 candidates were obtained.

\begin{figure*}[htbp]
\subfigure{
	\begin{minipage}[b]{.41\paperwidth}
		\includegraphics[width=9cm]{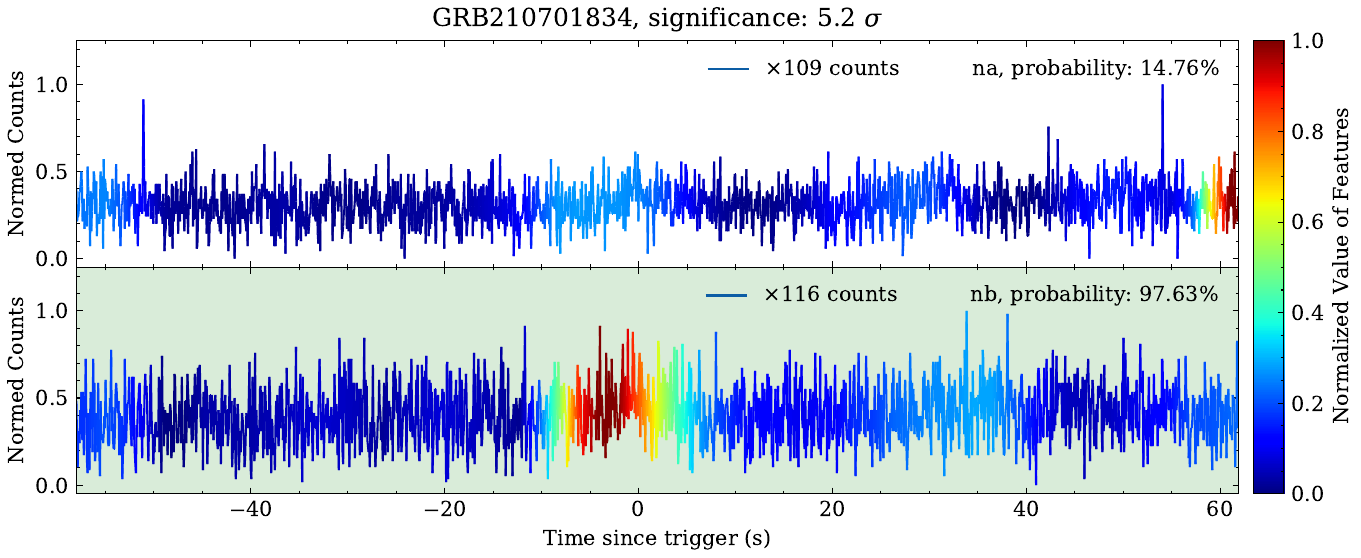}\vspace{7mm}
		\includegraphics[width=9cm]{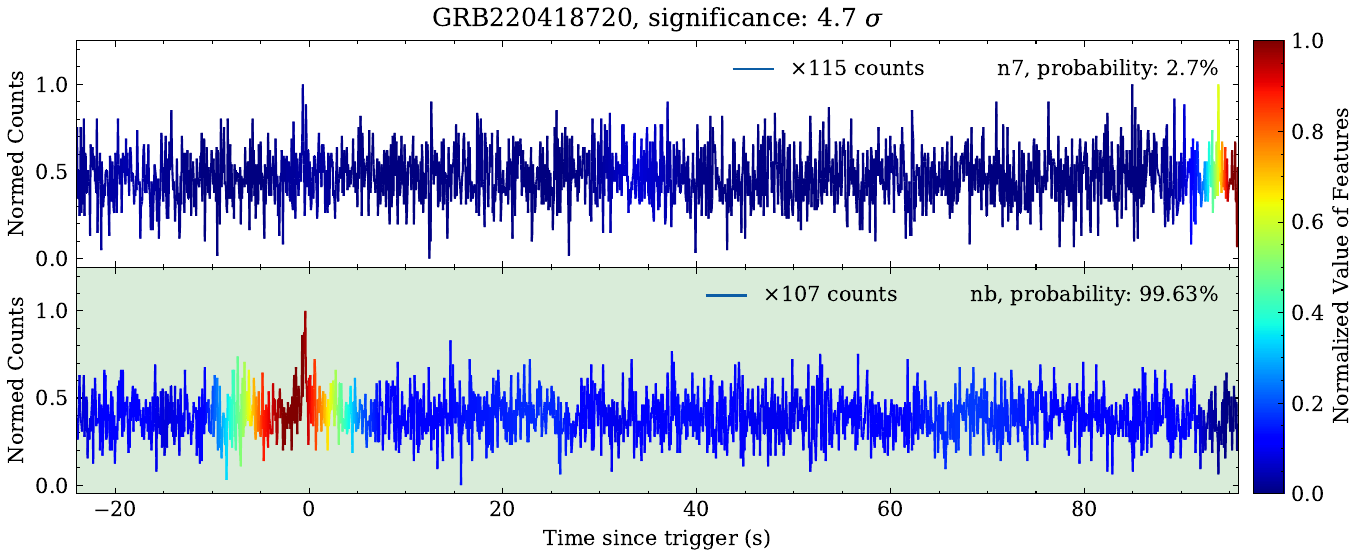}
	\end{minipage}
}
\subfigure{
	\begin{minipage}[b]{.41\paperwidth}
		\includegraphics[width=9cm]{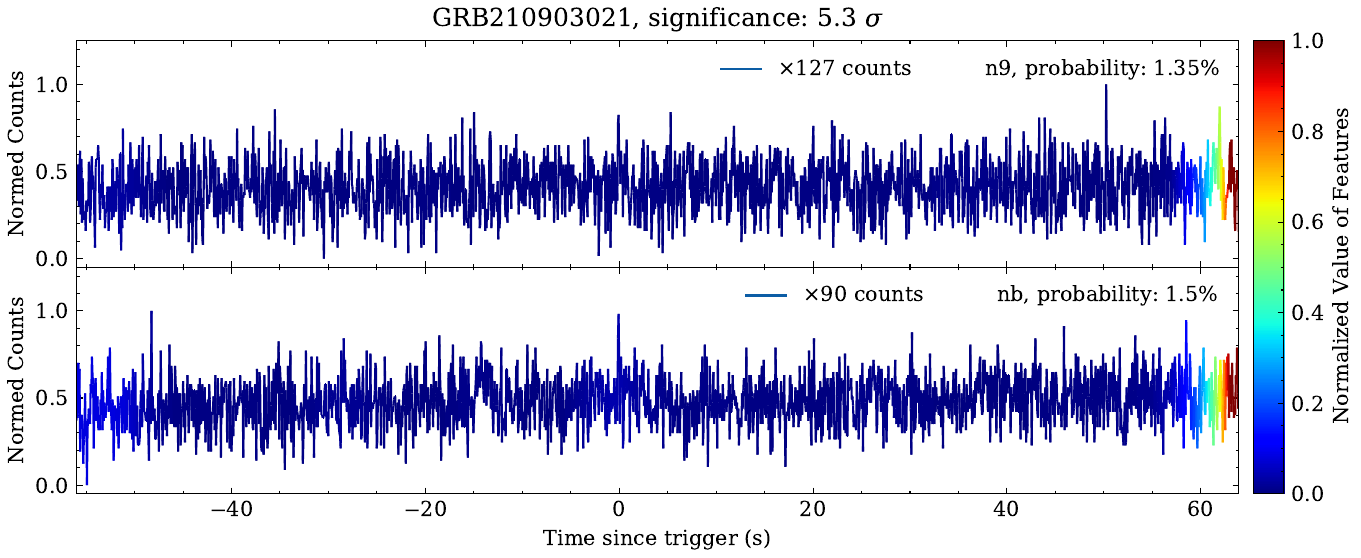}\vspace{7mm}
		\includegraphics[width=9cm]{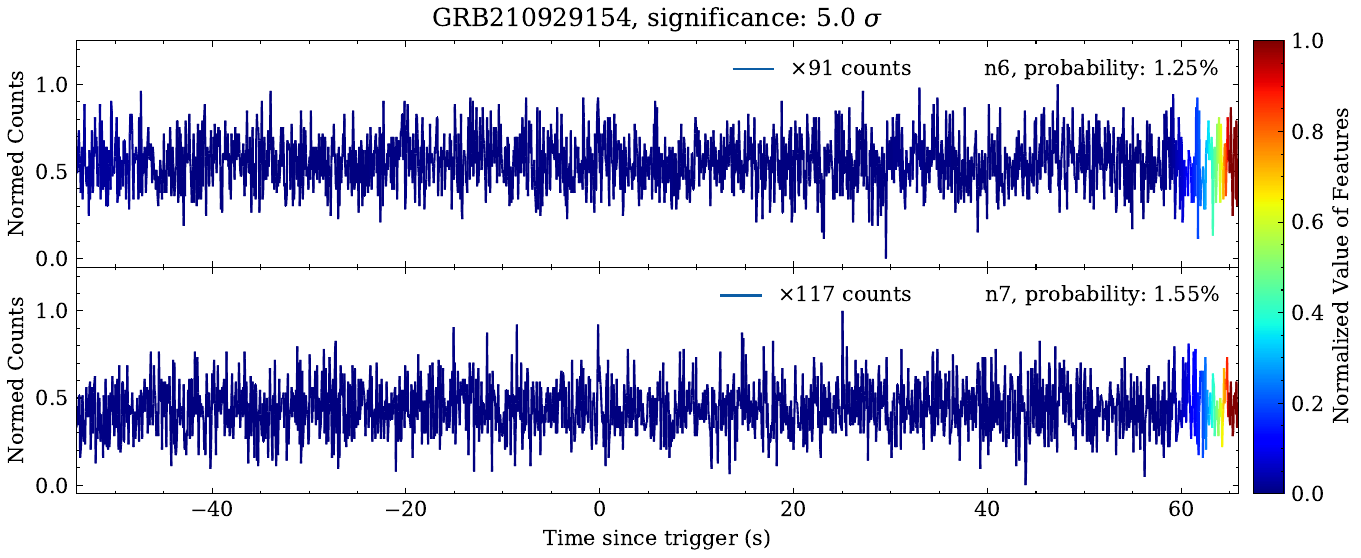}
	\end{minipage}
}
\caption{The mapping-curves of the 4 GRBs that our model failed 
to identify.
The green background indicates the model distinguished the burst 
signal. The time of the curve is relative to the trigger time of Fermi/GBM. 
The significance are derived from the TRIGDAT file of GBM burst data products.}
\label{fig:example_false_grb}
\end{figure*}

We matched these candidates with the catalog of Fermi/GBM triggers\footnote{\url{https://heasarc.gsfc.nasa.gov/W3Browse/fermi/fermigtrig.html}}, 
sub-threshold triggers\footnote{\url{https://gcn.gsfc.nasa.gov/gcn/fermi_gbm_subthresh_archive.html}} 
and sub-threshold GRBs.
The information of optimum matching is only about bursting time.
The result is shown in Table \ref{table:searched_event}. 
Approximately  98\% public GRBs of GBM instrument are matched with our candidates. 
The mapping-curves of features of the 8 missing GRBs are shown in Figure \ref{fig:example_false_grb}. 
We performed a checkback with our initial events, and find three of them are identified as 
single initial events within 120\,s time window. 
They seem like single detector triggered events. 
For the other 4 missed GRBs, we found that they are classified as non-GRB category with causes unknown. 
For the missed GRB 211031175, 
there are two corresponded initial events, but the time interval of the most significant 
features is longer than 20 seconds. Moreover, the significance of most missed GRBs are barely 
about 5$\sim$6\,$\sigma$, which may also be the reason for missing. 
There are still a large number of candidates 
that correspond to other triggering types, e.g., solar flare (SFLARE), soft gamma-ray repeater (SGR), 
local particles (LOCLPAR), etc. 
Morphologically speaking, those type triggers look like GRB events from their 
features of light curves and spectrum considerably. 
In general, the spatial locations of those triggers are apparently different from those of GRBs.
For our binary classification method, the training process and data sets do not include 
the location information of bursts. 
Therefore, we lack a way to screen candidates based on location information.

\begin{figure*}[htbp]
\hspace{1cm}
\subfigure{
	\begin{minipage}[b]{.4\paperwidth}
		\includegraphics[width=7cm]{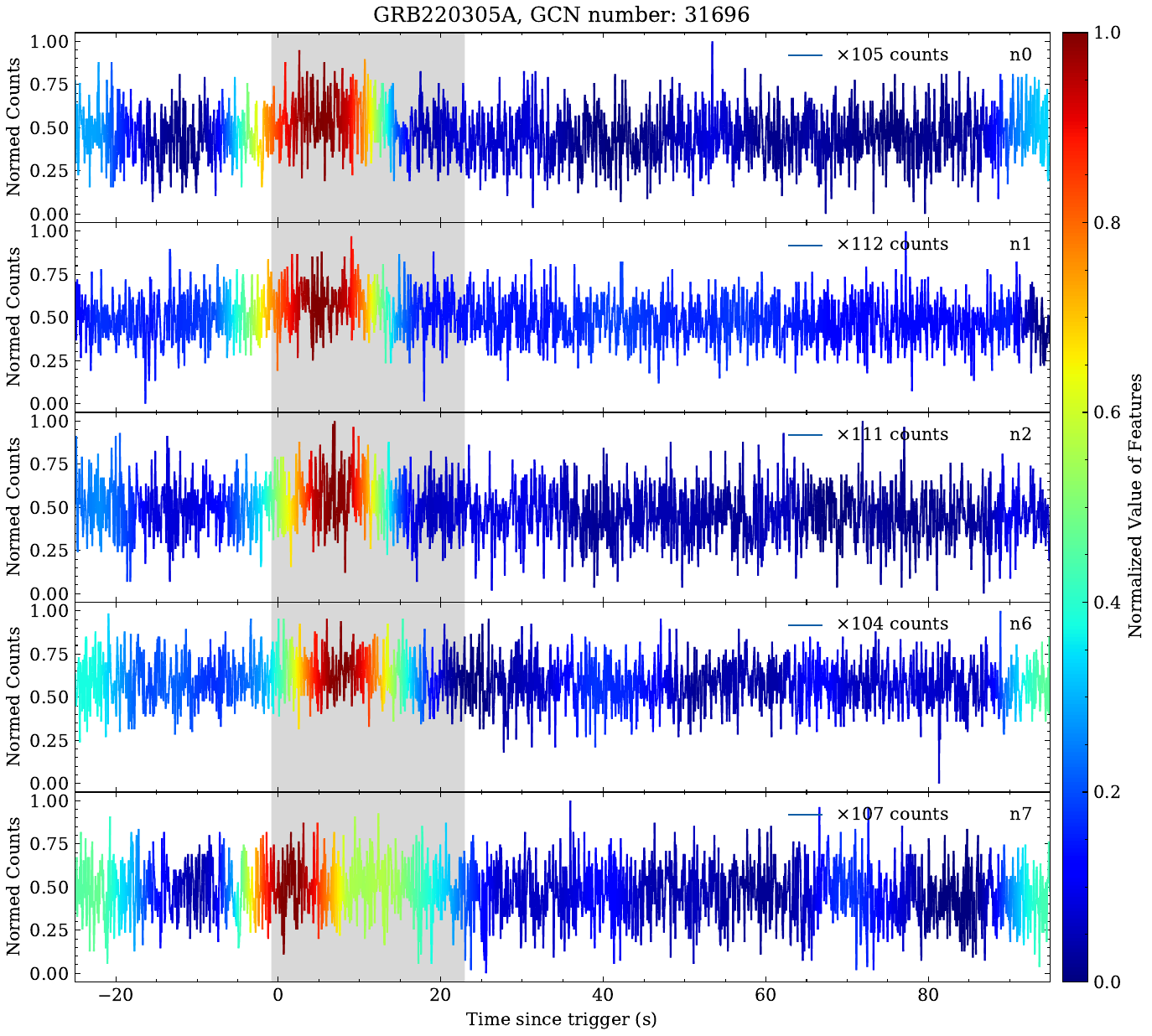}
		\includegraphics[width=7cm]{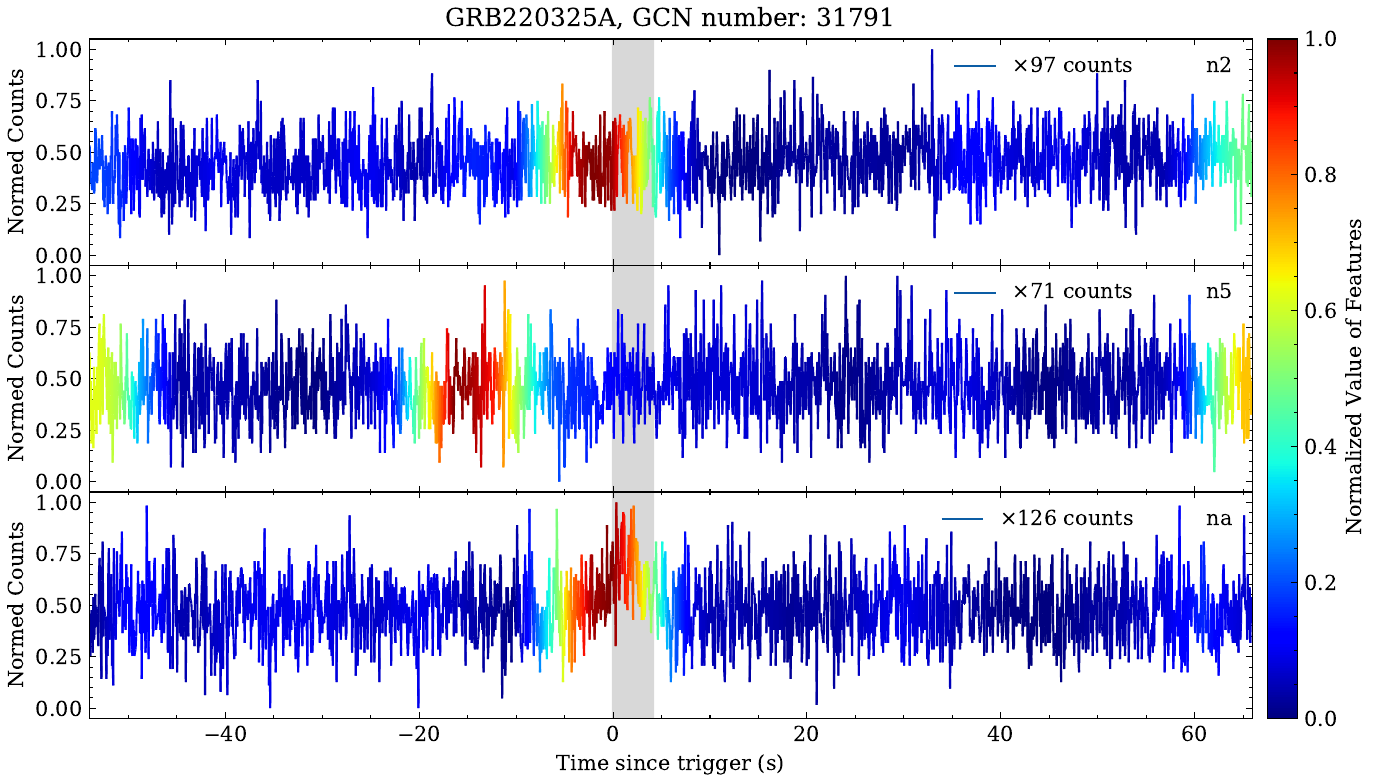}\vspace{0.75mm}
	\end{minipage}
} 
\subfigure{
	\begin{minipage}[b]{.4\paperwidth}
		\includegraphics[width=7cm]{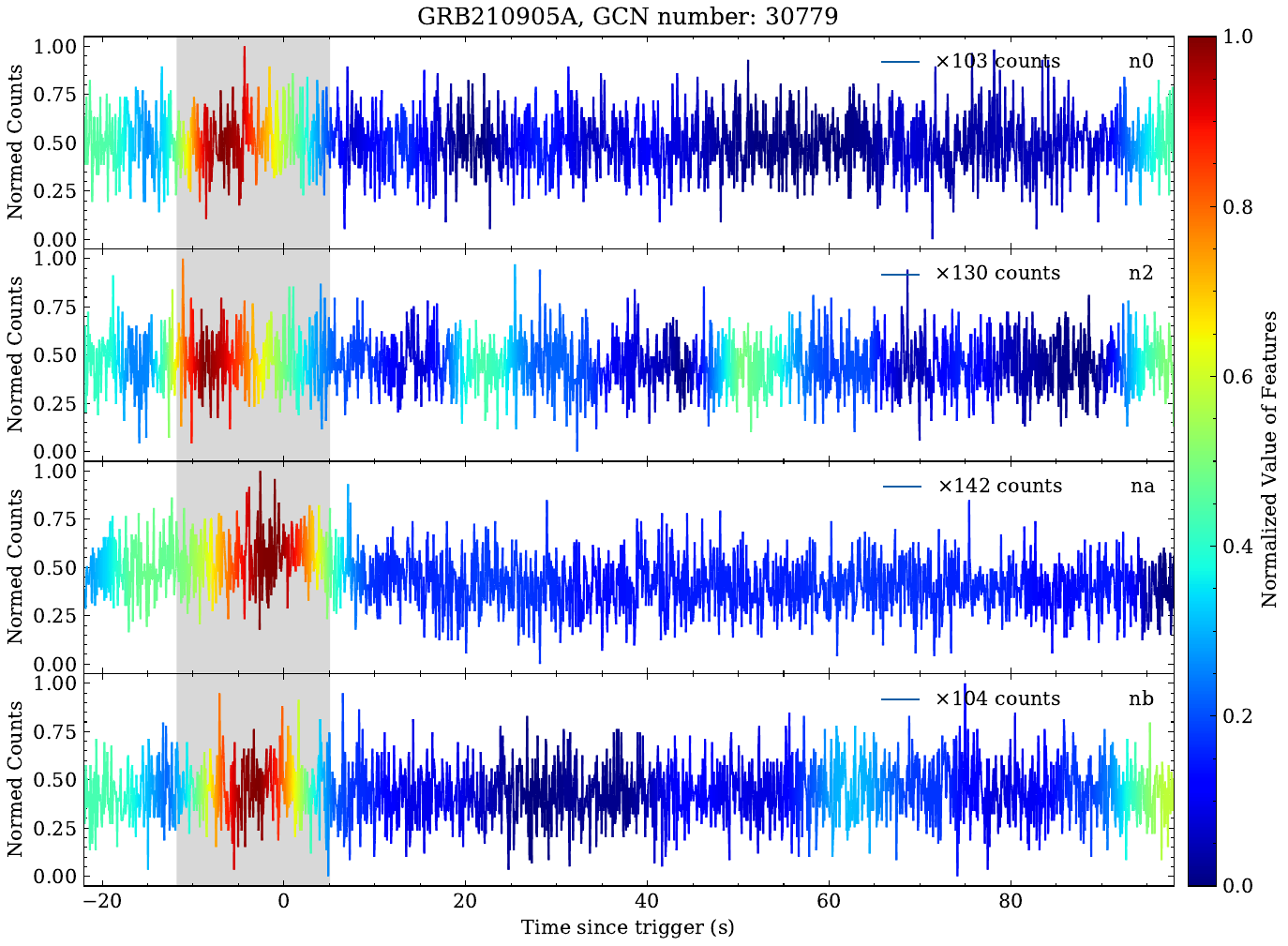}\vspace{0.75mm}
		\includegraphics[width=7cm]{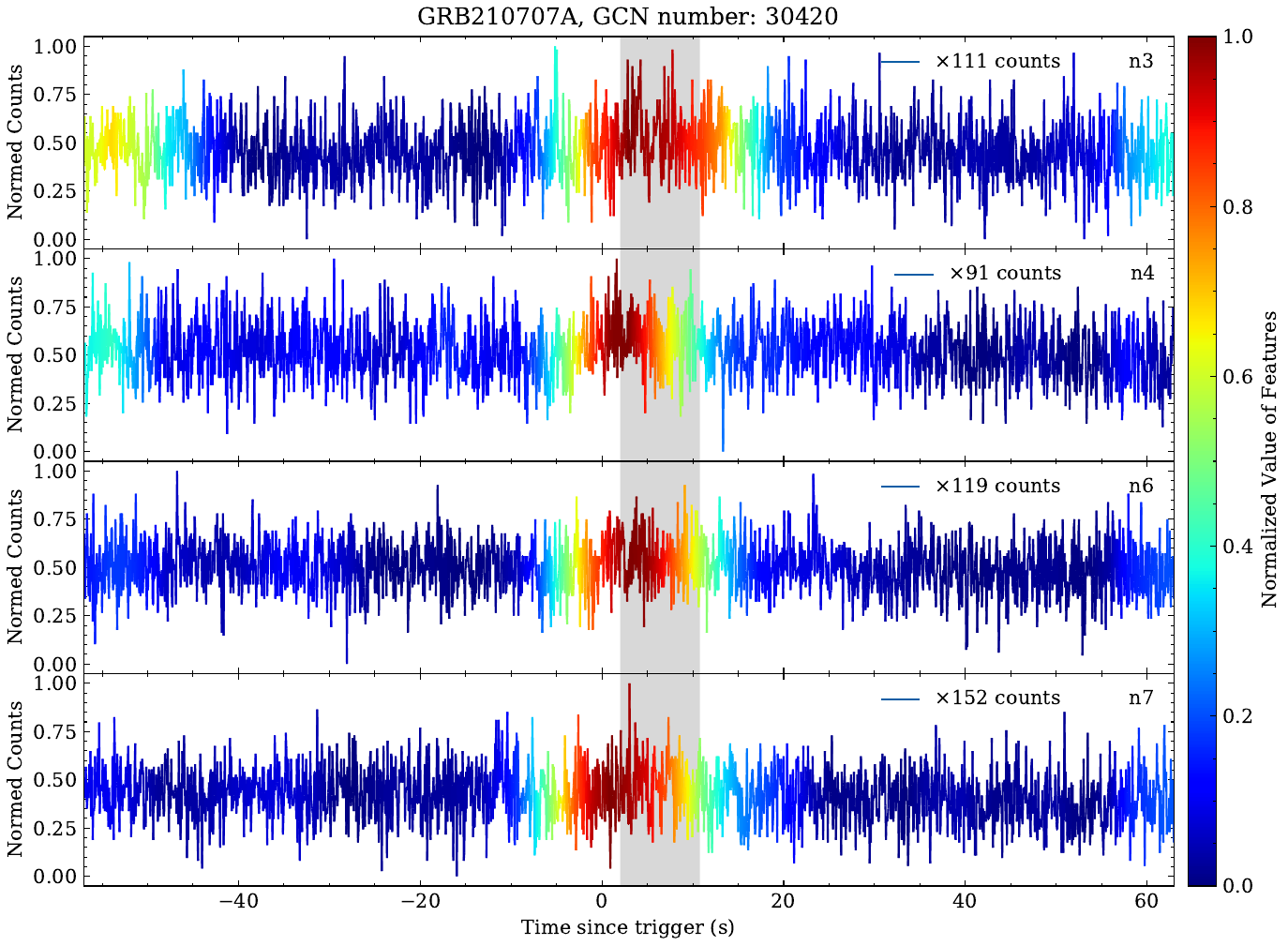}
	\end{minipage}
}
\par
\hspace{1cm}
\subfigure{
	\begin{minipage}[b]{.4\paperwidth}
		\includegraphics[width=7cm]{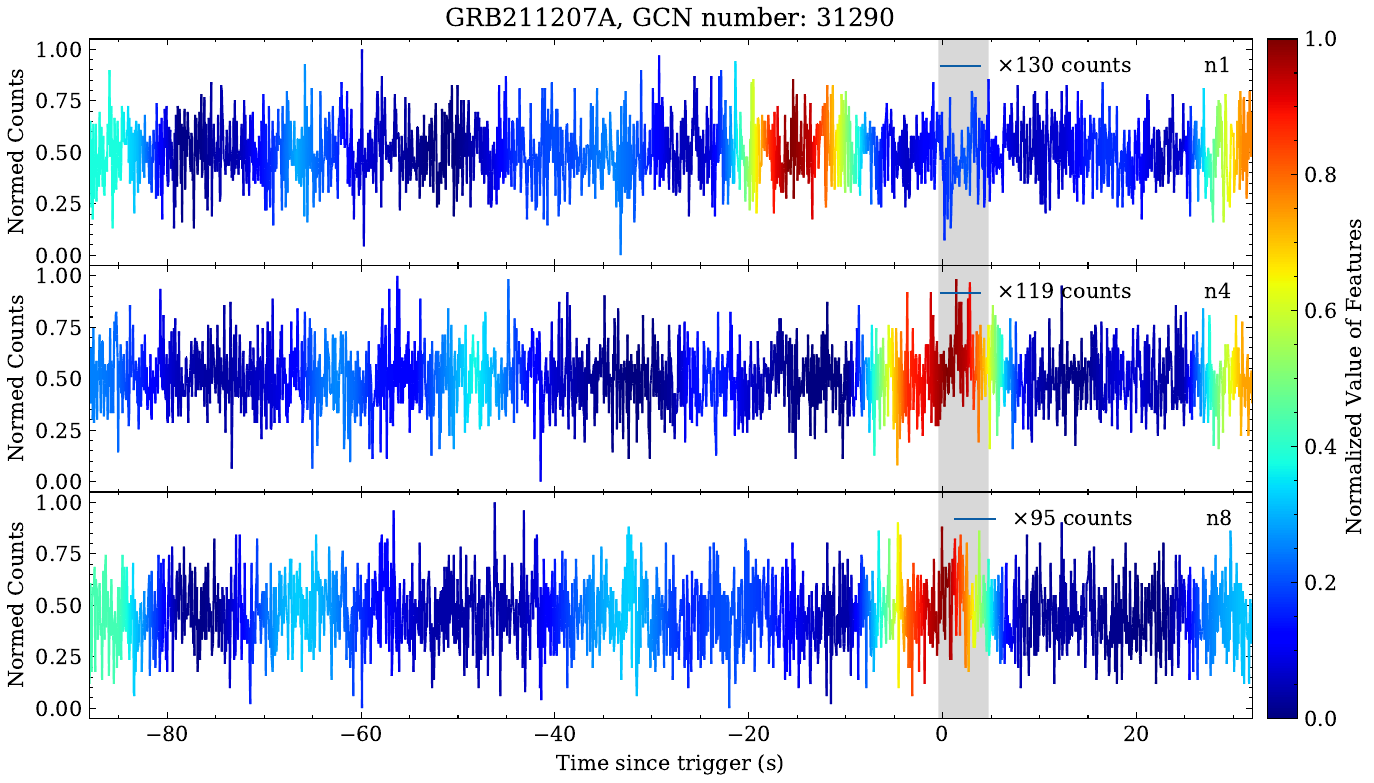}
	\end{minipage}
}
\subfigure{
	\begin{minipage}[b]{.4\paperwidth}
		\includegraphics[width=7cm]{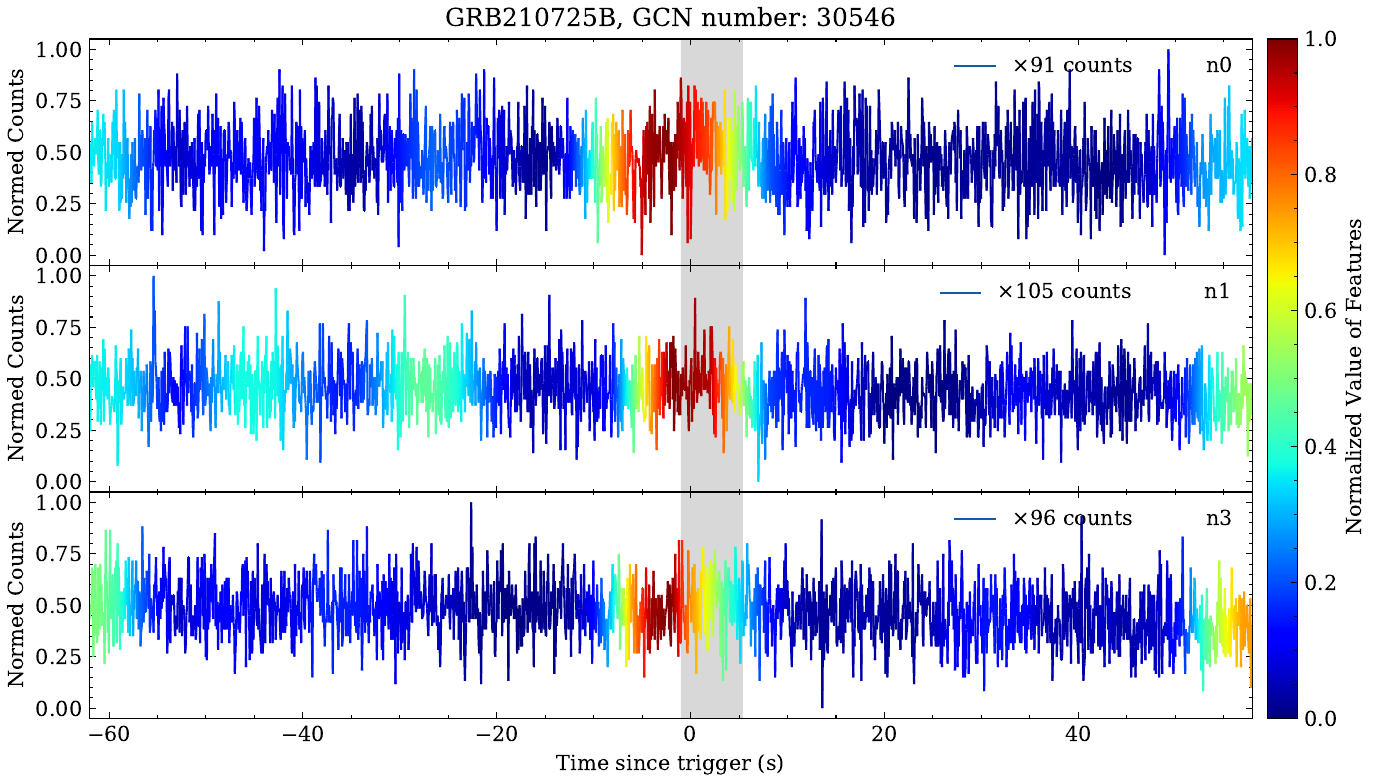}
	\end{minipage}
}
\par
\hspace{1cm}
\subfigure{
	\begin{minipage}[b]{.4\paperwidth}
		\includegraphics[width=7cm]{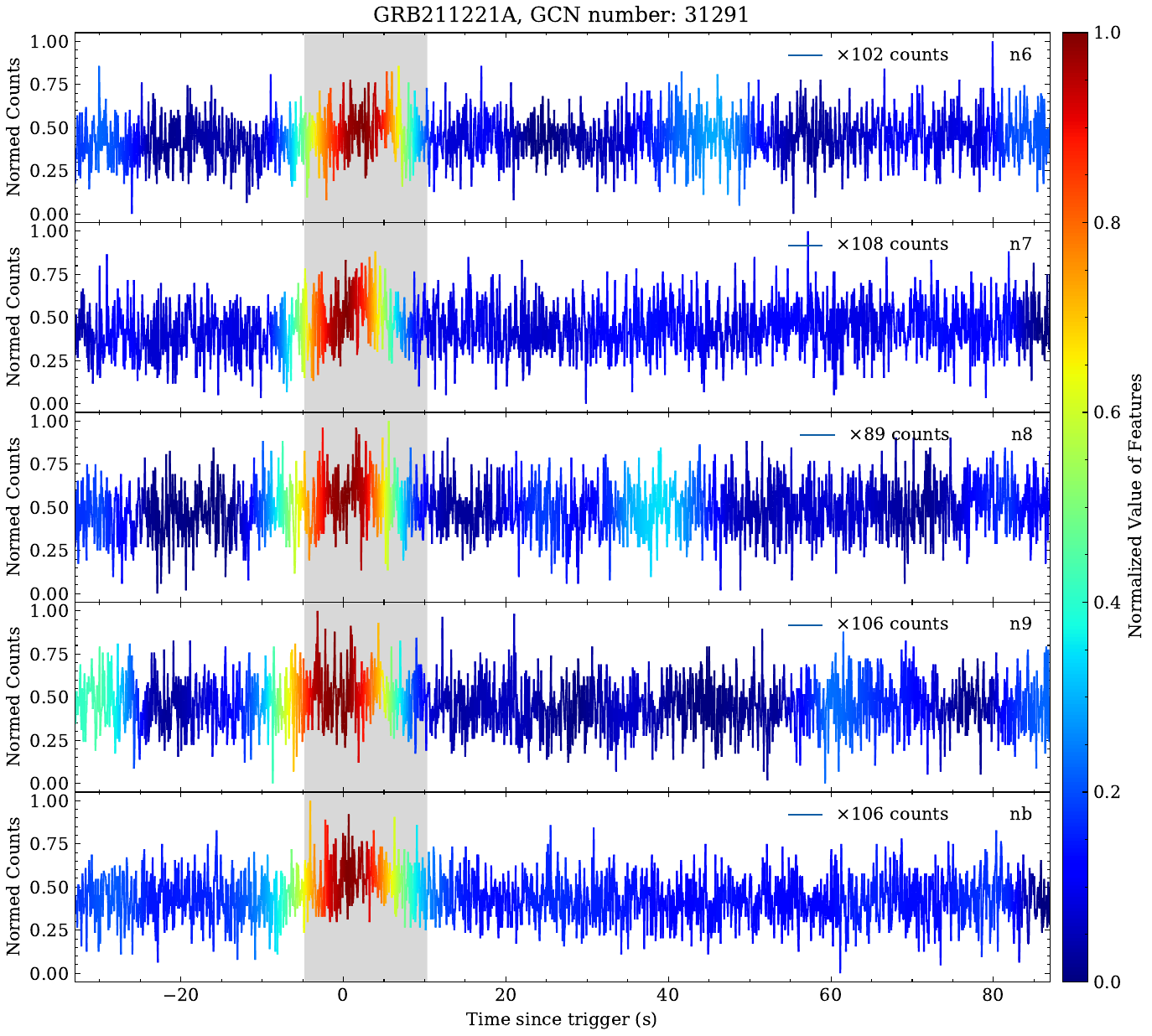}
	\end{minipage}
} 
\subfigure{
	\begin{minipage}[b]{.4\paperwidth}
		\includegraphics[width=7cm]{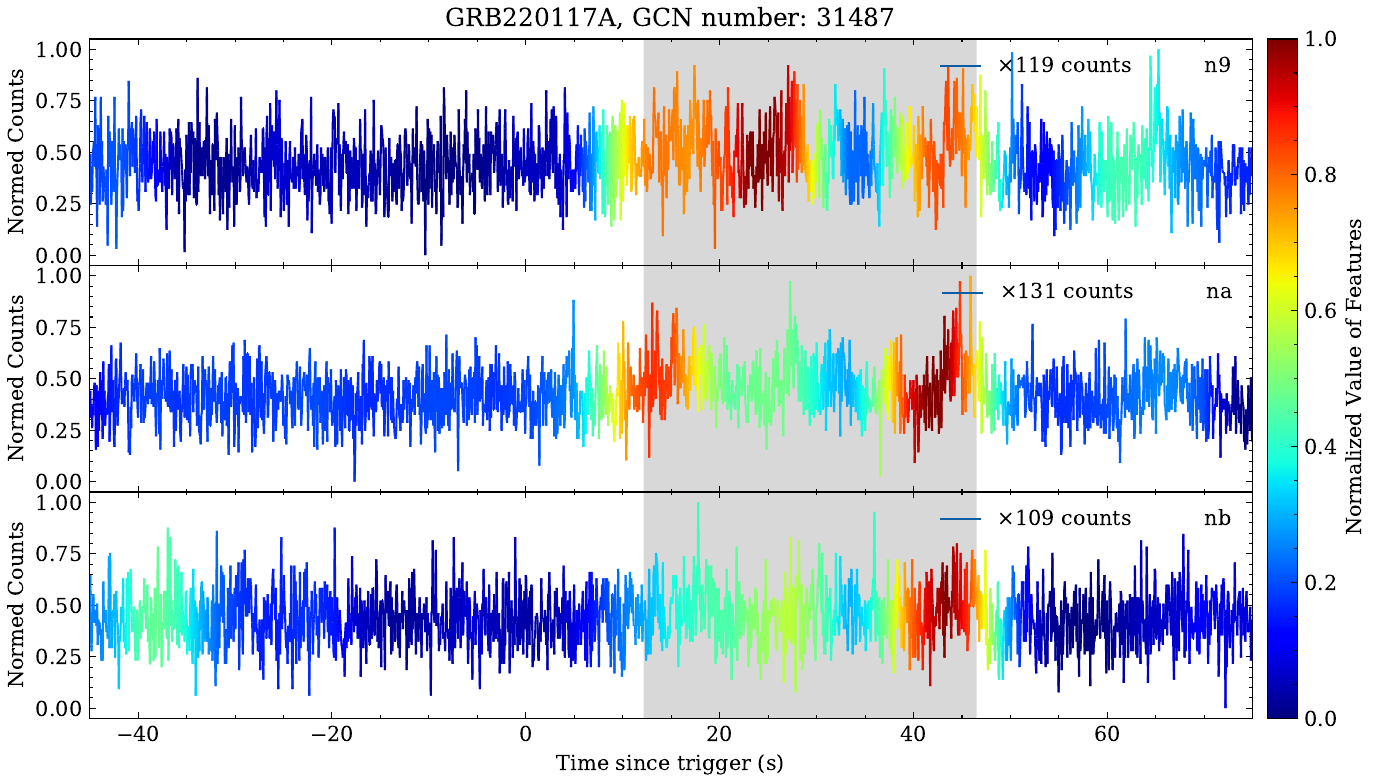}\vspace{0.75mm}
		\includegraphics[width=7cm]{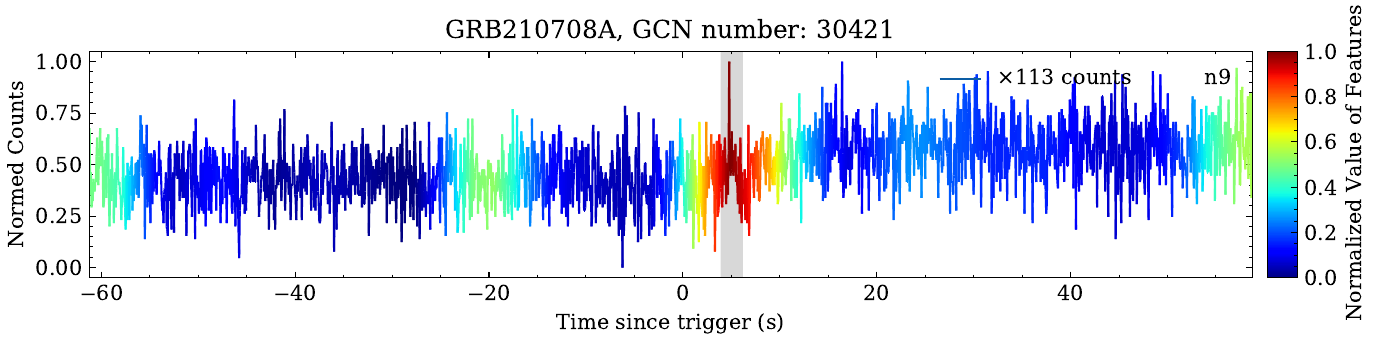}
	\end{minipage}
}
\caption{The mapping-curves of seven sub-threshold GRBs (GCN circular) 
that our model identified. The symbol $\ast$ indicates that the model distinguished the burst 
signal on only one detector. 
The time of the curve is relative to the trigger time in the GCN. 
The gray period is $T_{90}$, and its calculation method is shown in the Appendix \ref{sec:appendix_cal_snr_t90}.
}
\label{fig:gcn_sub_grb}
\end{figure*}

In addition, we further compared our candidates with published sub-threshold GRBs 
and sub-threshold triggers of Fermi/GBM. 
There were 11 sub-threshold GRBs published on the Gamma-ray Coordinates 
Network (GCN) circular\footnote{\url{https://gcn.gsfc.nasa.gov/gcn3_archive_GRB.html}} totally in this year. 
Eight sub-threshold GRBs therein were found to be temporally coherent with our candidate. 
Our model classified them as GRB category with high confidence, whose mapping-curves are 
displayed in Figure \ref{fig:gcn_sub_grb}. 
Their characteristics performed by our model are prominent and accord with analysis in GCN. 
One sub-threshold GRB (with GCN number 30421) is individually comparable to an initial event 
that shows high confidence in one detector only. 
The remaining two GRBs were not successfully identified by our model. 
The result demonstrates that our model exhibits excellent discriminative 
ability for low-threshold gamma-ray burst signals. 
Furthermore, we conducted a comparative analysis between our candidates and the 
sub-threshold triggers that may potentially include weak GRBs. 
With the screening conditions we set, 64 sub-threshold triggers therein were effectively distinguished. 
The search accuracy rate is a bit low, about 9.96\%. 
Our manual inspection reveal 120 initial events that still have extremely high confidence level.
This suggests that by loosening the screening conditions, these events might be treated as GRB category. 
The identification accuracy would increase to 28.66\%.

By comparing GRBs in Fermi burst catalog with 
GRBweb\footnote{\url{https://user-web.icecube.wisc.edu/~grbweb_public/Summary_table.html}}, 
we find additional 51 GRBs are not listed in Fermi burst catalog but were detected by other instruments.
Among them, there are 17 GRBs without corresponding GBM continuous data. 
Fermi/GBM has observed the other 34 GRBs normally during their burst phase. 
In other words, GBM burst advocates (BA) did not ascertain these GRBs via ground analysis. 
Amazingly, 21 of the 34 GRBs are temporally matched with our unknown events discovered by our model.
Among them, 5 GRBs show significant signals in multiple 
detectors, 16 GRBs only display in single detector, and the other 13 GRBs do not have 
burst signal in the corresponding count maps.  
The GRB names, relevant GCN numbers, instruments and the analysis results of our discovered GRBs are 
listed in Table \ref{table:gbm_not_grb_list}. 
Interestingly, we find the trigger time of two GRBs, GRB220308A and GRB220403C (from GRBweb), 
are close together with GRB220308233 and GRB220403424 that listed in the Fermi/GBM catalog, respectively. 
In our opinion, these two GRBs could be peculiar long bursts with precursor component comprised long interval 
comparable with GRB160625B \citep{GRB_160625B} and GRB221009A \citep{GRB_221009A}.

\begin{figure*}[htbp] 
\centering 
\includegraphics[width=15cm]{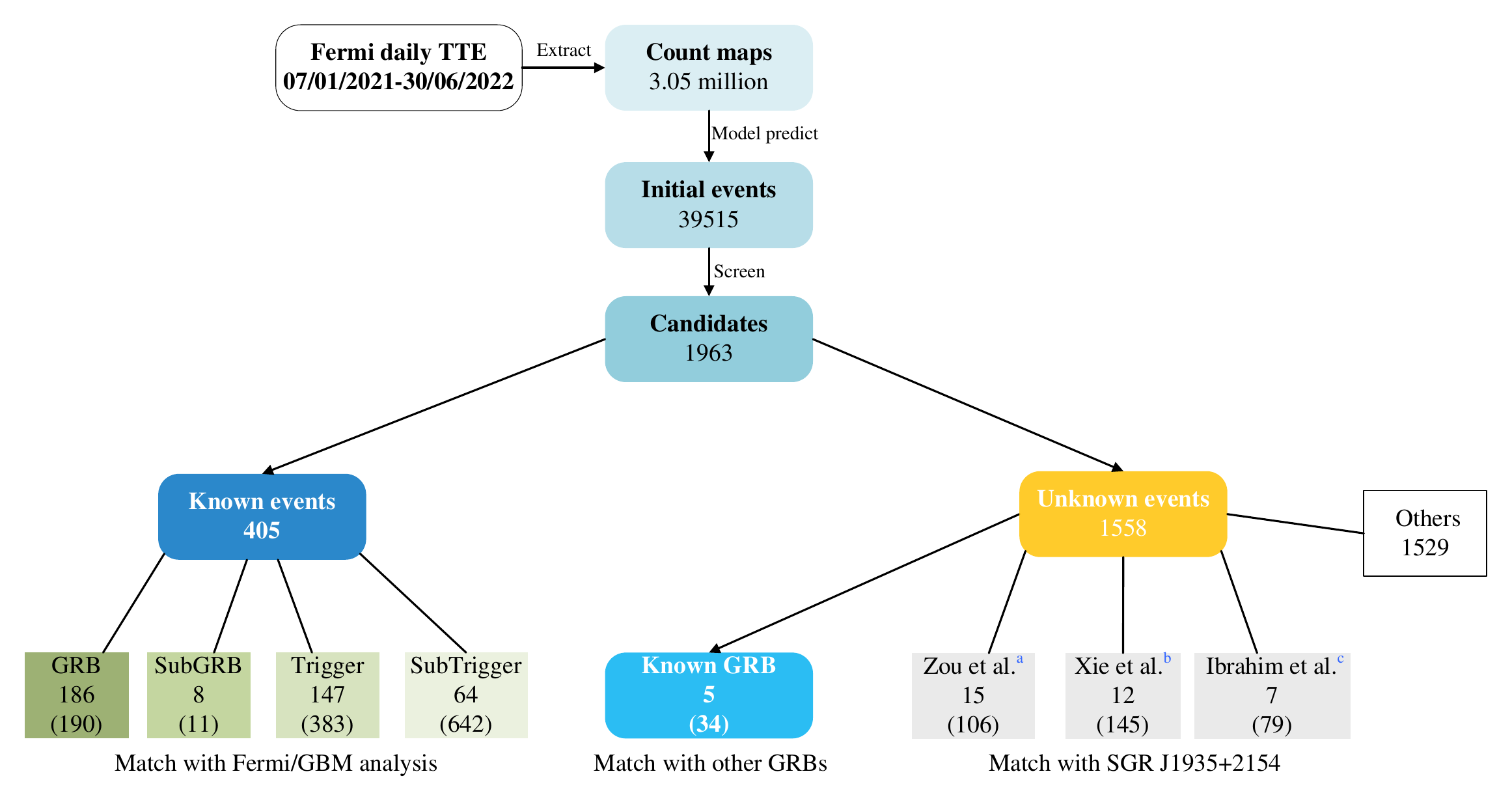} 
\caption{The process of our searching method, and the matching result.
We compared our 1963 candidates with events that existed in Fermi catalogs and literatures, 
\footnotesize{$^a$ \cite{sgr_list_zoujinhang}, }
\footnotesize{$^b$ \cite{sgr_list_xieshenglun}, }
\footnotesize{$^c$ \cite{sgr_list_Rehan}}. 
The number in parentheses present the published bursts in total. 
There are 1536 candidates left, shown as 'Other', which indicates the nonmatched events.
These results are corresponding to Table \ref{table:searched_event} and 
Table \ref{table:candi_sgr_match}. 
}
\label{fig:candi_category}
\end{figure*}

The Figure \ref{fig:candi_category} illustrates our process of searching for candidates 
and the results that compared with known events. 
For verifying the scientific significance of these 1558 unknown events, 
we could compare them with other published intense bursts in this year, e.g., 
magnetars, solar flares. Here we only choose well known magnetar SGR J1935+2154 as a target. 
By using the sub-threshold search algorithm, \citet{sgr_list_linlin2}, \citet{sgr_list_linlin}, 
\citet{sgr_list_zoujinhang}, \citet{sgr_list_Rehan} and \citet{sgr_list_xieshenglun} have found 
over 200 new bursts from SGR J1935+2154 than Fermi trigger catalog. 
And it is important to note that there are 74 duplicate bursts in these references after our verification.  
Then, we compared these new bursts with our unknown events. 
24 of the 1558 unknown events corresponded to these newly discovered bursts. 
The contrasted results are shown in Table \ref{table:candi_sgr_match}, 
and their nameplate (ID), relevant researches, and our analysis results are listed in Table \ref{table:candi_sgr_match_list}. 
This indicates that our model has the capability to identify weak bursts comparable to manual analysis.
It means our unknown events could potentially comprise some non-negligible scientific events. 
Besides, the other 1529 unknown events are respectively shown in 
Table \ref{table:candi_snr_greater_5} (SNR $ \ge 5\,\sigma$), 
Table \ref{table:candi_snr_less_5} (SNR $ \textless 5\,\sigma$)
and Table \ref{table:candi_snr_no} (without SNR) according to their SNR information. 
The $T_{90}$ and SNR are calculated with reference to Appendix \ref{sec:appendix_cal_snr_t90}.

\begin{figure*}[htbp]
\hspace{-1.3cm}
\subfigure[ ]{
	\begin{minipage}[b]{.47\paperwidth}
		\includegraphics[width=9cm]{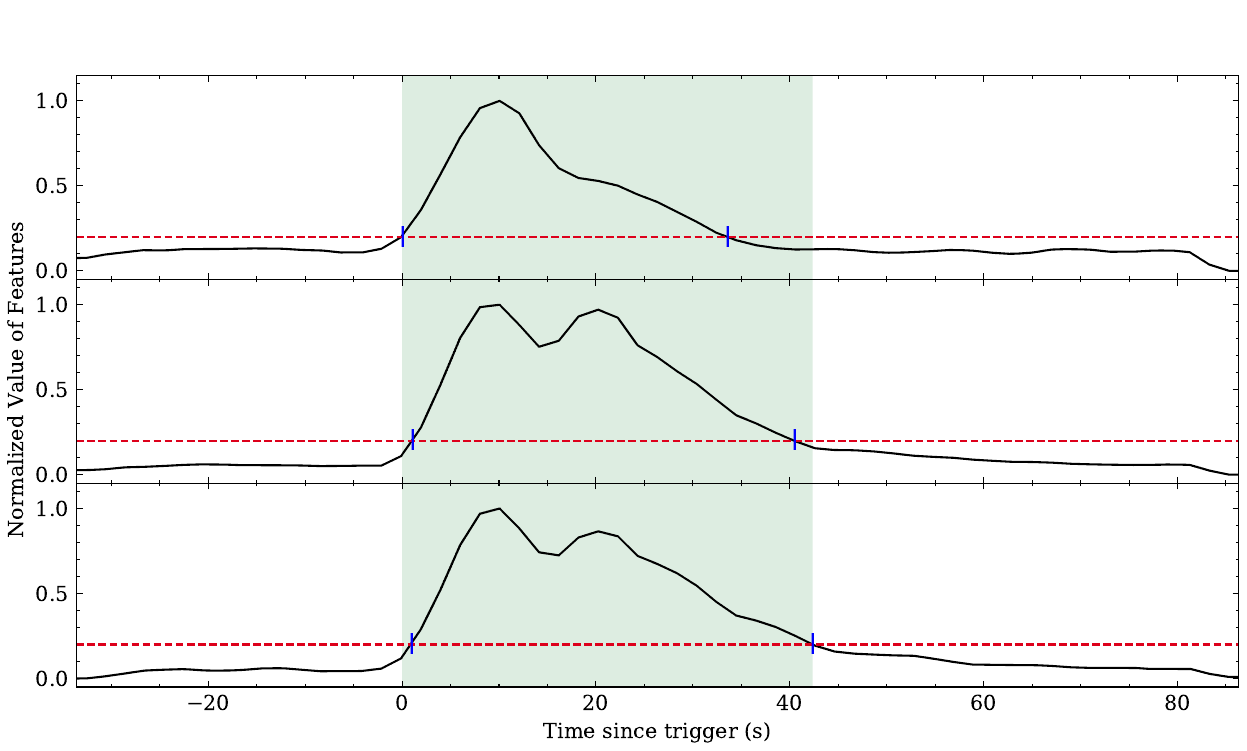}
	\end{minipage}
}
\subfigure[ ]{
	\begin{minipage}[b]{.47\paperwidth}
	    \includegraphics[width=10cm]{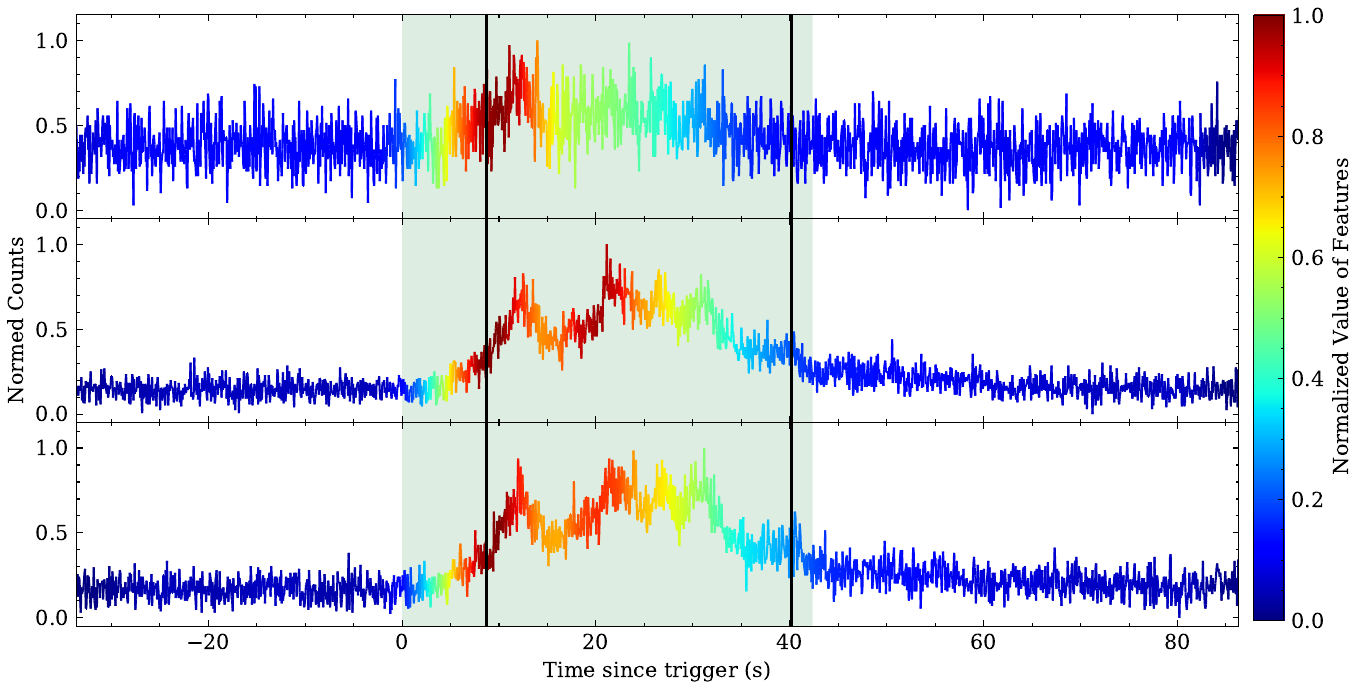}
	\end{minipage}
}
\par
\hspace{-1.3cm}
\subfigure[ ]{
	\begin{minipage}[b]{.47\paperwidth}
		\includegraphics[width=9cm]{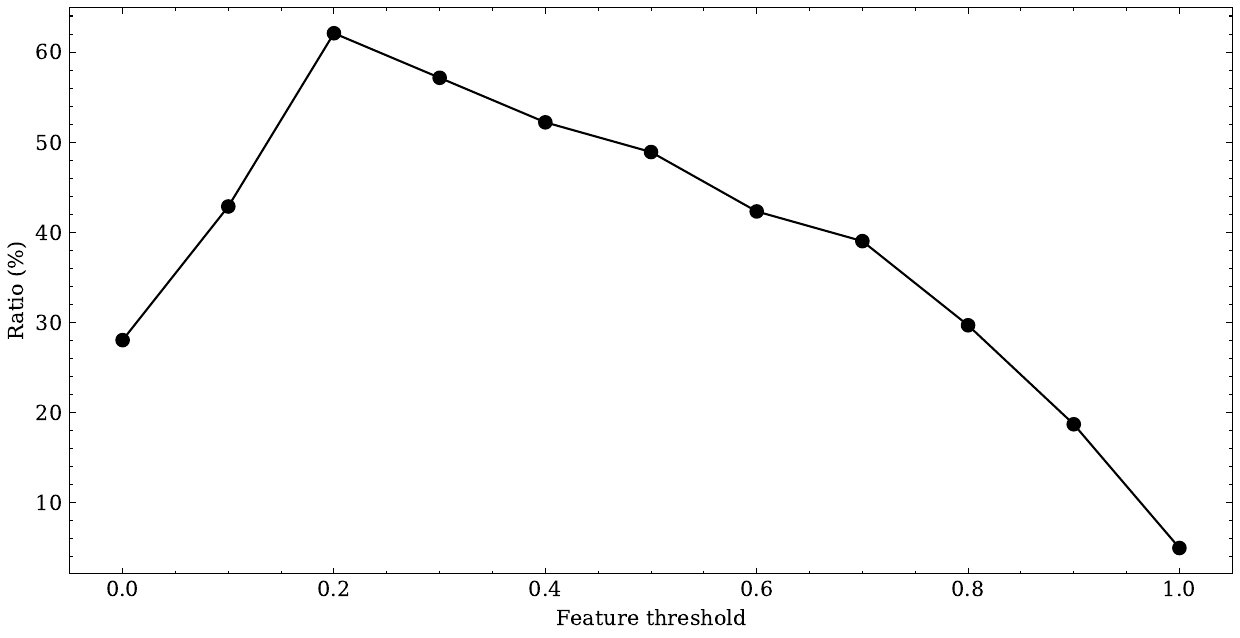}
	\end{minipage}
}
\subfigure[ ]{
	\begin{minipage}[b]{.47\paperwidth}
		\includegraphics[width=10cm]{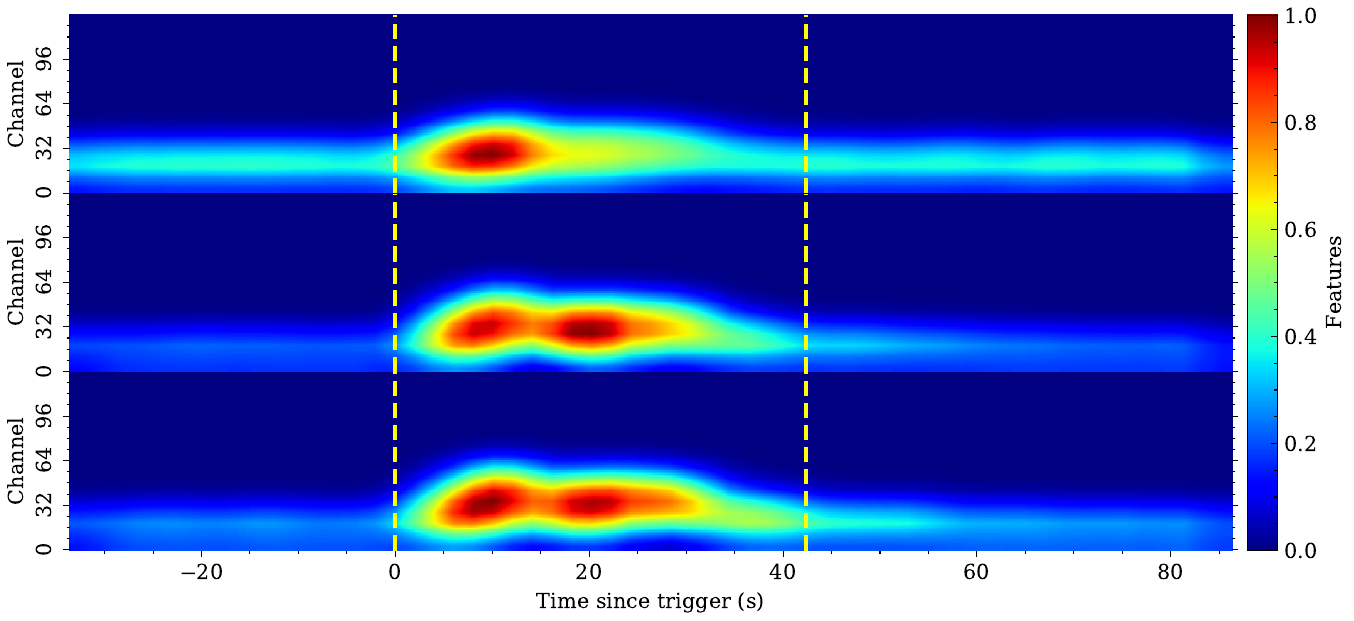}
	\end{minipage}
}
\caption{Right: the heat-map of feature and the mapping-curves of feature. 
Left: curve obtained by summing the heat-map of feature 
along channel dimension and the IoU (Fermi-$T_{90}$ and $T_{90, F}$) are 
$>$ 0.3 with various feature thresholds 
from the identified GRBs in Section \ref{sec:application}. 
The red dashed lines are the set feature threshold. 
The green area represents the period of $T_{90, F}$. 
The black vertical line indicates the Fermi-$T_{90}$. 
}
\label{fig:cal_f_t90}
\end{figure*}

The information of duration and location are necessary to analysis our searched candidates. 
However, our examination demonstrate that most of the candidates are too weak for such calculation. 
The prominent areas in heat-maps are relevantly corresponded with the burst region of GRBs, 
hence we can use the period of prominent area to characterize the duration of the candidate. 
Employing the mapping-curves of feature (see Figure \ref{fig:cal_f_t90}b) for each candidate 
and normalizing the summed value of feature per bin, 
the corresponding feature curves are finally formed, see Figure \ref{fig:cal_f_t90}a. 
The duration of each initial event $T_F$ is determined as the time period of the prominent area 
exceeds the threshold. By verifying of the duration $T_{90}$ and $T_F$ of some GRB samples, 
we establish a correlation between Intersection over Union (IoU, always $\leq 1$) ratio, 
as shown in Figure \ref{fig:cal_f_t90}c. 
The threshold is empirically selected as 0.2 and $T_F$ of initial events for each candidate is summed as its duration $T_{90, F}$, 
shown as the green area of Figure \ref{fig:cal_f_t90}b. 
We use $T_{90, F}$ to characterize the duration of bursts/candidates. 
The heat-map of this example is marked as $T_{90, F}$, see Figure \ref{fig:cal_f_t90}d, 
which distinctly demonstrates the close connection between prominent area and $T_{90, F}$.

\begin{figure*}[htbp] 
\centering 
\includegraphics[width=15cm]{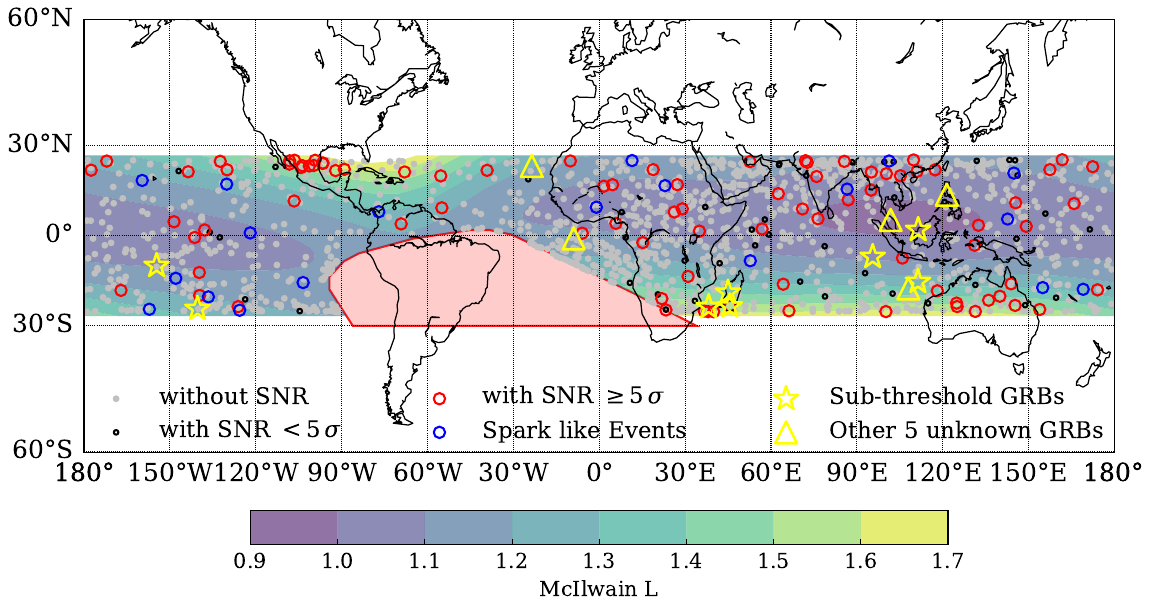} 
\caption{Location of the Fermi spacecraft in orbit at the $T_{90, F}$ 
start time of the candidates. 
The color gradient shows the geomagnetic latitude according to the McIlwain L. 
If the spacecraft is in the region where the McIlwain L is greater than 1.5, 
the candidate is likely to be associated with local particles. 
}
\label{fig:candi_earth_point}
\end{figure*}

\begin{figure*}[htbp] 
\centering 
\includegraphics[width=15cm]{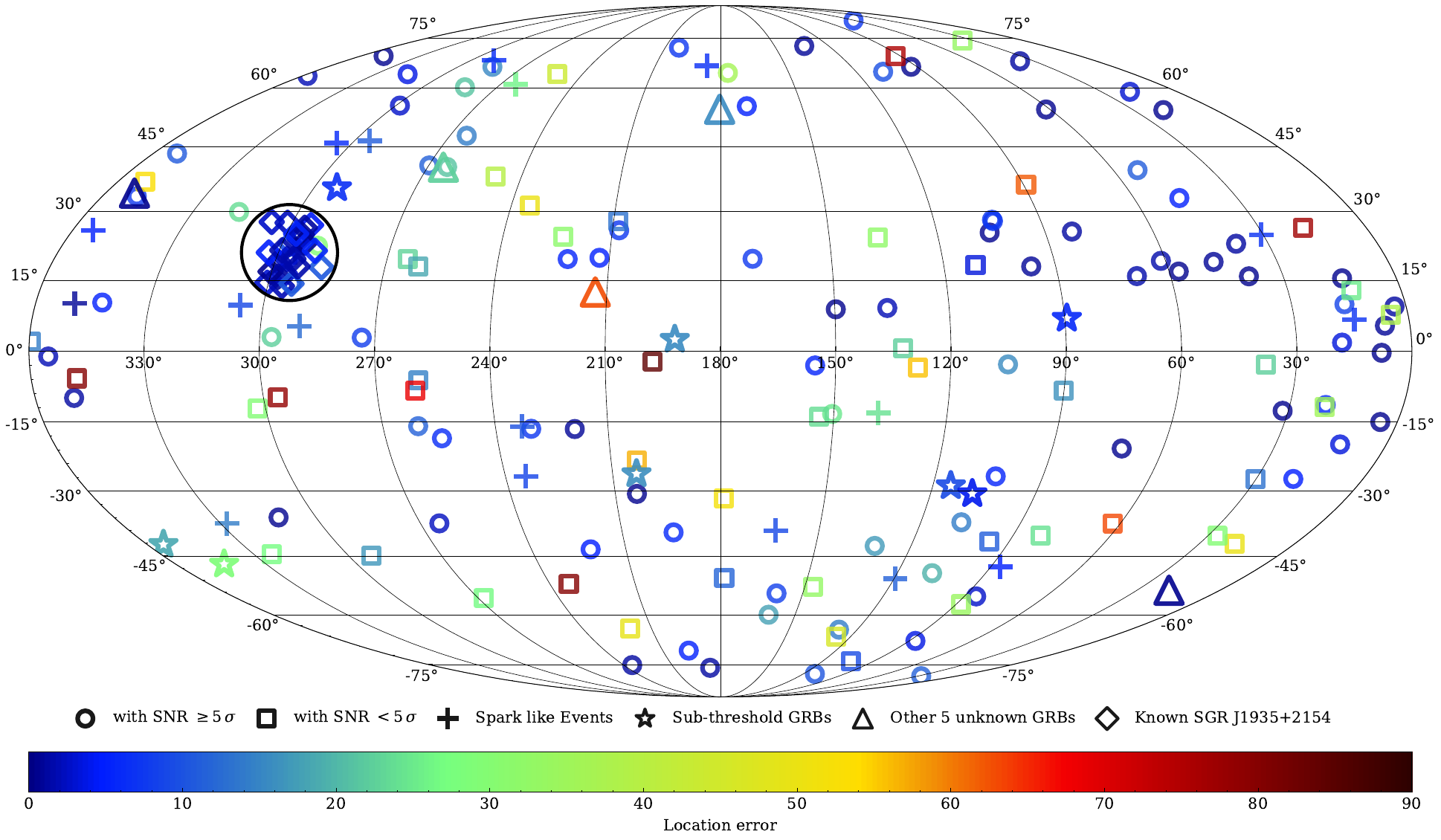} 
\caption{The localization of the unknown events. 
The color bar illustrates the localization error of events within 1 $\sigma$ 
equaled to the radius of spatial position error.
The thick black circle encloses the events from SGR J1935+2154. 
}
\label{fig:candi_loc}
\end{figure*}

The $T_{90, F}$ of all 1529 unknown candidates are computed and their start times 
are shown in Table \ref{table:candi_snr_greater_5}, Table \ref{table:candi_snr_less_5} 
and Table \ref{table:candi_snr_no}, respectively. 
Based on the start time of each unknown candidate, we obtained its corresponding locations of Fermi 
satellite in orbit, shown in Figure \ref{fig:candi_earth_point}. 
The candidates with different SNR are displayed. 
Some candidates show significant signals only in a single bin, thus we label them as spark like events. 
Our spark like events, sub-threshold GRBs, and other 5 known GRBs are also marked.
Considering the correlation between 
certain signal, e.g., particle events (LOCLPAR), TGF (Terrestrial Gamma-ray Flashes), 
with geomagnetic latitude (McIlwain L), 
we also present McIlwain L in the figure.
We find there is no obvious distribution or relationship between our candidates and McIlwain L.
The localization results of the candidates in these tables are calculated based on 
the algorithm in the Appendix \ref{sec:appendix_locate}. 
The localization of these 181 unknown events, with their $T_{90}$ valid computing, 
were calculated and shown in Figure \ref{fig:candi_loc}. 
The localization error of these candidates are tightly related to their significance. 
The candidates with SNR $ \ge 5\,\sigma$ including spark like events present accurate localization.
From these unknown candidates, there are 4 events located in the approaching region of SGR J1935+2154, 
and their light curves show simple peak shape that similar to burst of SGR. 
We also sign 24 candidates corresponded to SGR J1935+2154 mentioned above. 
We suggest those 4 events are probable candidates of SGR J1935+2154. 
There are 41 unknown candidates located in earth occlusion region, and 
15 candidates would direct to sun through their localization. 
Referring localization information, the angular distribution of these localized candidates is  
fairly uniform, which reveal the potential cosmic signals they are.

\section{Discussion and Conclusion}
\label{sec:discussion} 
As the research hotspots of multi-messenger and time-domain astronomy in particular, 
under the limitations of the existing data sets, it is a hot pursuit for researchers to 
making rational use of GRBs light curves and spectra to realize rapid and accurate 
discovery of their pertinent and meaningful information.
Therefore, deep learning methods have been gradually applied 
and have achieved great success in terms of identifying, classifying, and forecasting burst events 
and phenomena of astronomy. 
In this paper, the supervised deep learning methods were applied to identify GRBs 
from the Fermi/GBM TTE data. 
Referring to the burst catalog of GBM, we categorized GRBs and non-GRBs sets and achieved  
the aim of training valid models. 
Then we applied the optimal model to identify GRB candidates from the approaching one-year TTE data.

\par
Thousands of count maps of GRBs and non-GRBs were used, and the valid training set, 
validation set, and test set were built, see Table \ref{table:data set} for detail.
The selected 2560 GRBs were multiplied by the number of triggered detectors, 
and they were taken as the samples of GRB category. These samples  
were distributed into data sets in three time period. Correspondingly, daily data products 
after filtering out the burst TTE data were randomly selected as samples of non-GRB category. 
Our sample construction method is conducive to minimizing the limitation of insufficient positive sample size as much as possible.
Moreover, the reasonable ratio of each dataset is advantageous to ensure the balance 
of data distribution. 
Furthermore, the time period distributing of samples would avoid data confusion. 
There are three types datasets that contain samples with bin size of 256, 128, and 64\,ms, respectively, 
and the purpose is to consider the effect of time resolution. 
A total energy of 128 channels were chosen  
for all samples. Therefore, the pixel matrix of count maps of samples is 128*1875, 128*938 and 128*469. 

\par
There are three single network models structured in our work, named as 
\textit{plain}-CNN, ResNet and ResNet-CBAM, respectively, shown in Figure \ref{fig:network_architectures}. 
Base on the general Conv-block, we progressively added residual modules, residual-attention modules to the \textit{plain}-CNN model, and appeared to be working well. 
The three models are trained on three data sets with different time bins separately. 
We find that the ResNet-CBAM model achieved the highest accuracy on the 64\,ms data set. 
This model incorporates the attention mechanism of spatial and channel dimensions, as proposed by \citet{model_block_CBAM}, 
to significantly enhance the feature extraction ability, which is crucial for improving classification. 
The training curves of models in Figure \ref{fig:learning_curve} show tendency of overfitting, 
even though we employed the data expansion, dropout and early stopping mechanisms. 
Subsequently, collection of burst events, data augmentation, and model hyper-parameter optimization may solve this problem. 
Relatively, as shown clearly in Table \ref{table:network metrics} and Figure \ref{fig:confusion_matrix}, 
models with same architecture that trained on data sets of smaller time bin shown better performance. 
The models achieve high accuracy in the training set, validation set and test set. 
The model's comparable accuracy on the validation and test sets indicates its strong generalization capability. 
Intuitively the ROC curves in Figure \ref{fig:model_ROC} have indicated the powerful classification ability 
of ResNet-CBAM model. It is possible that count map with shorter time bins contains more 
fine-grained information of bursts, as it is generally agreed that the variation of some burst structure 
behave differently within different time bin \citep{relate_diff_bin_explain}. 
Different model architectures exhibit varied performance as due to their feature extracting capability. 
We tried to fuse those single network models through four combinations. Overall of fusion architecture is 
shown in Figure \ref{fig:network_fusion}. Four fusion algorithms with new FC layers are structured and 
trained with three different bin size data sets with their performance shown in Table \ref{table:network_fusion}. 
Fusion models improve accuracy indeed, however, they consume more computing resources and the improvement is not impressive yet.

\par
\begin{figure*}[htbp]
\hspace{-1.3cm}
\subfigure{
	\begin{minipage}{.47\paperwidth}
		\includegraphics[width=10cm]{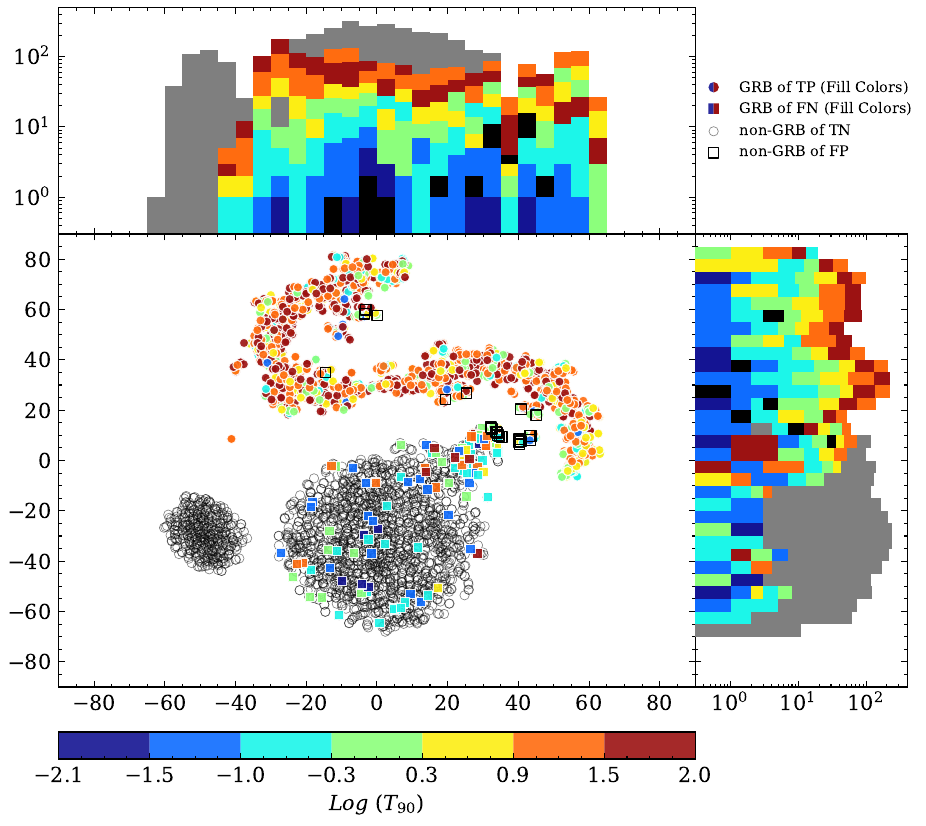}
	\end{minipage}
}
\subfigure{
	\begin{minipage}{.47\paperwidth}
		\includegraphics[width=10cm]{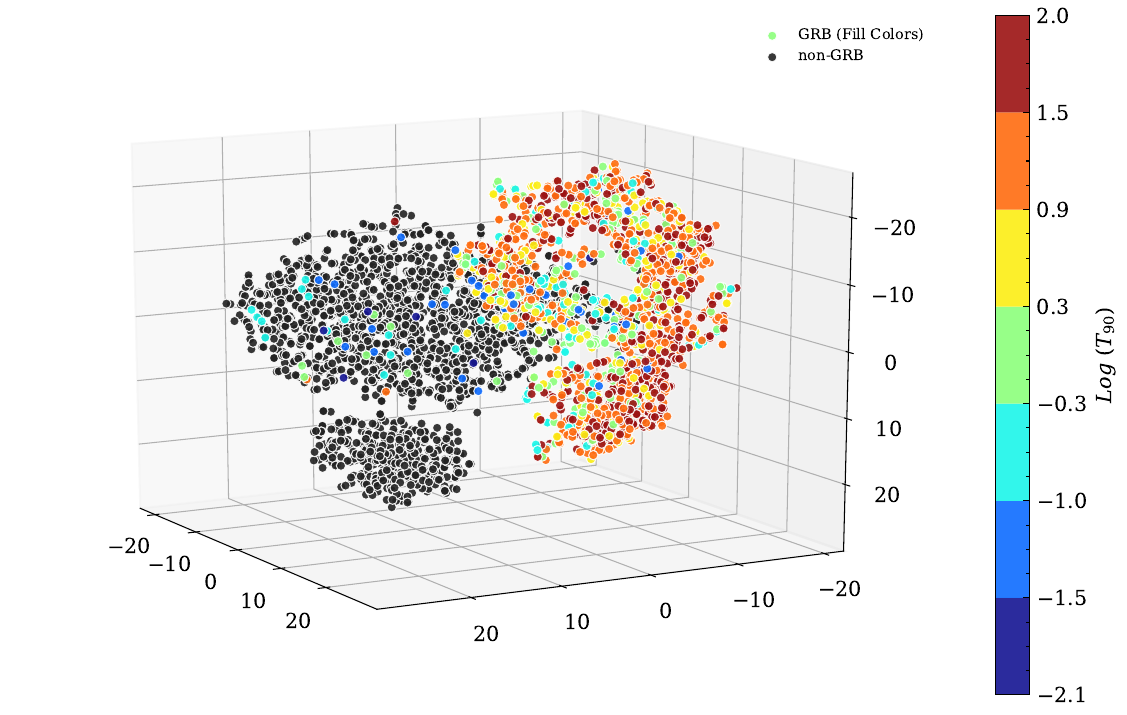}
	\end{minipage}
}
\caption{
The 2D and 3D visualization of the feature maps is achieved using the 
dimensionality reduction method t-SNE, with $T_{90}$ embedded for each sample. 
The feature maps are output by the last convolutional layer of optimal 
model on test set. 
The distributions of feature maps in feature space for GRB and non-GRB category 
are indistinguishable.
The feature maps of GRB and non-GRB categories show a clear aggregations. 
The duration of the GRB of FN samples was relatively short. 
}
\label{fig:feature_testset_tsne_t90_2d_3d}
\end{figure*}
\begin{figure*}[htbp]
\hspace{-1.3cm}
\subfigure{
	\begin{minipage}{.47\paperwidth}
		\includegraphics[width=10cm]{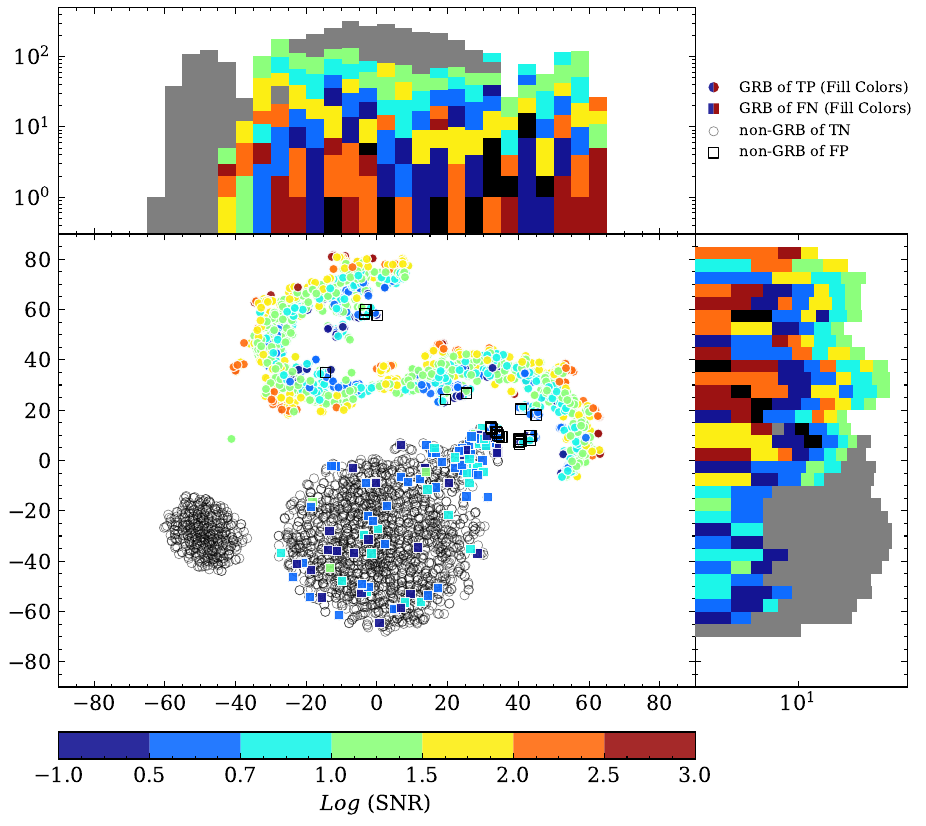}
	\end{minipage}
}
\subfigure{
	\begin{minipage}{.47\paperwidth}
		\includegraphics[width=10cm]{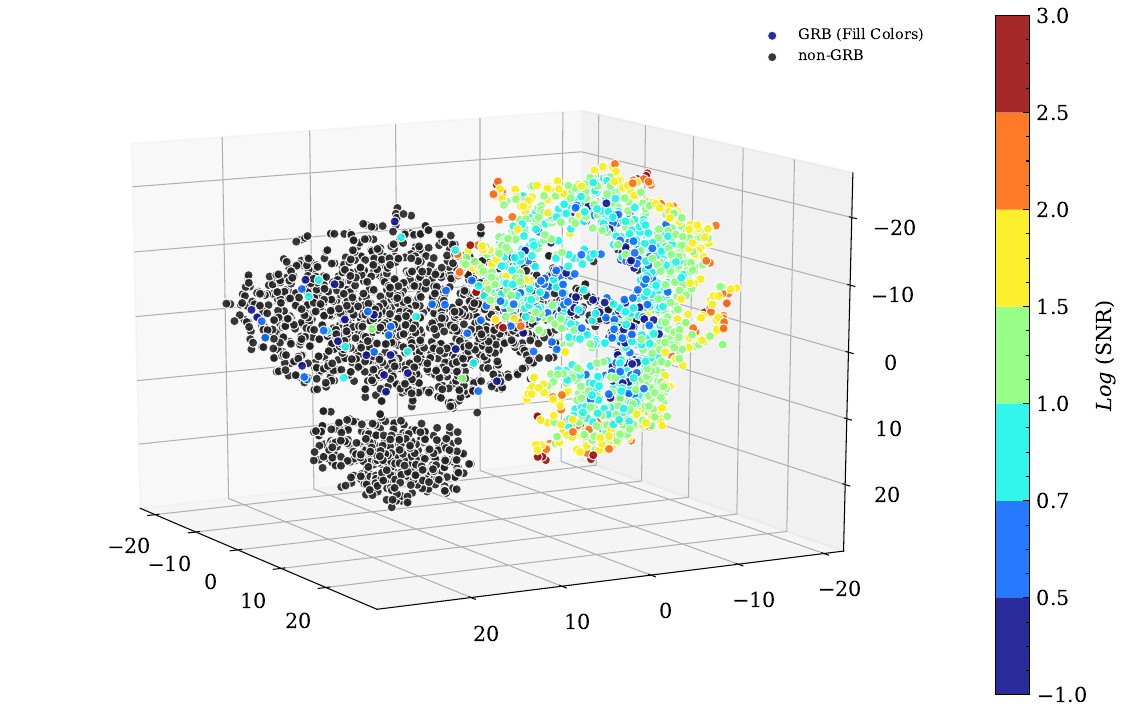}
	\end{minipage}
}
\caption{
The 2D and 3D visualization of the feature maps is achieved using the 
dimensionality reduction method t-SNE, with SNR (calculation see Appendix \ref{sec:appendix_cal_snr_t90}) embedded for each sample. 
The SNR of the GRB of FN samples was relatively low. 
}
\label{fig:feature_testset_tsne_snr_2d_3d}
\end{figure*}
\begin{figure*}[htbp]
\hspace{-1.3cm}
\subfigure{
	\begin{minipage}{.47\paperwidth}
		\includegraphics[width=10cm]{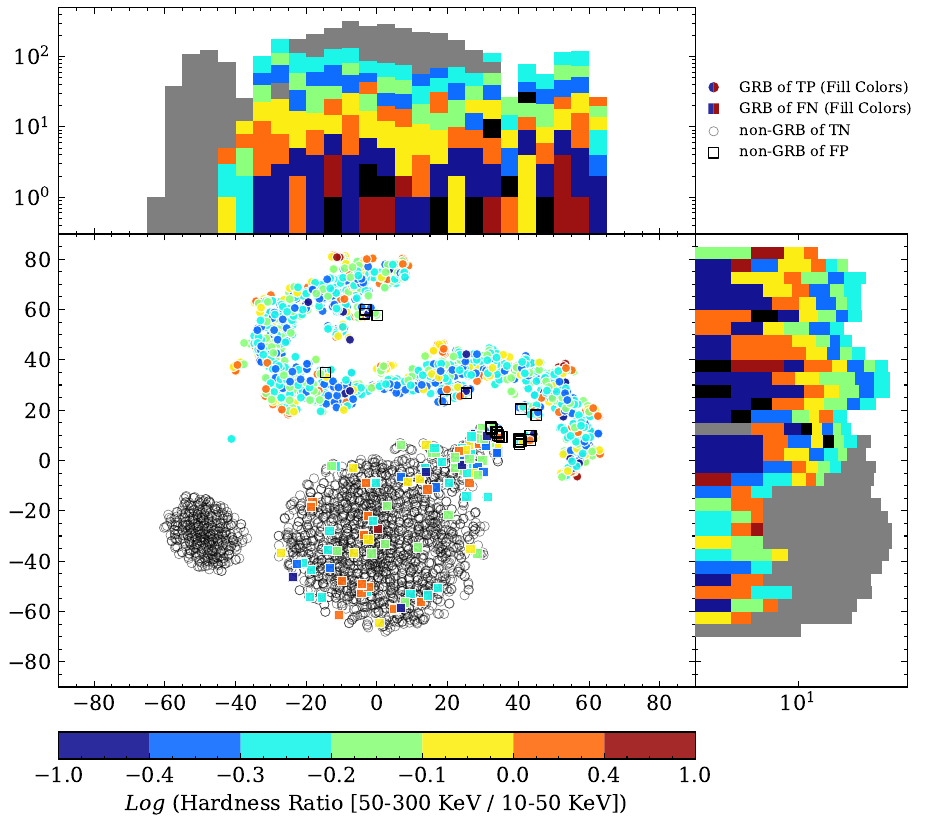}
	\end{minipage}
}
\subfigure{
	\begin{minipage}{.47\paperwidth}
		\includegraphics[width=10cm]{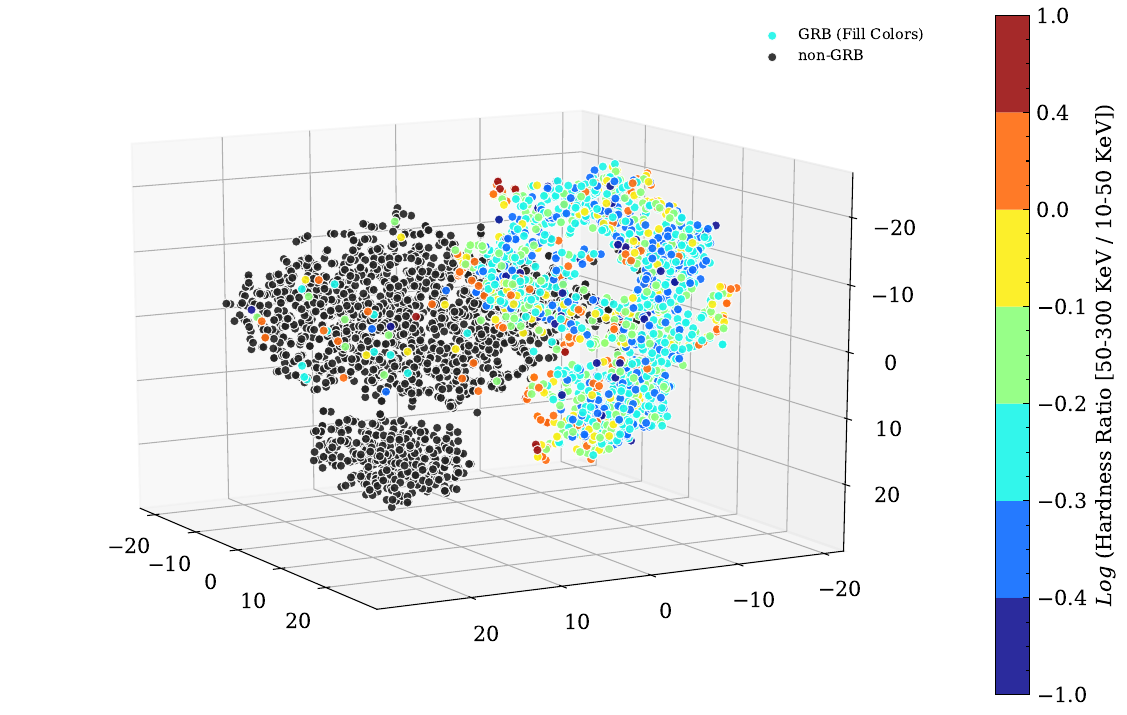}
	\end{minipage}
}
\caption{
The 2D and 3D visualization of the feature maps is achieved using the 
dimensionality reduction method t-SNE, with hardness ratio embedded for each sample. 
There is no clear pattern is found in the hardness ratio of the GRB of FN samples.
}
\label{fig:feature_testset_tsne_hardness_2d_3d}
\end{figure*}

In order to better understand the deep learning models, visualization analysis methods are employed. 
We use Grad-CAM approach to produce heat-map of features that extract by deep learning models. 
The bottom panels in Figure \ref{fig:lc_example} show the examples of the results which were extracted by 
the final convolutional layer of optimal model viz the ResNet-CBAM that trained on 64\,ms bin data set. 
From the mapping-curves of features we achieved, we could clearly 
find the magnitude of the bursts in these two feature figures. 
The t-SNE technique is also used for visualizing analysis. 
For comparison, the count maps and feature maps in test set, corresponding 
to unapplied and applied deep learning methods, are visualized as 2D and 3D diagrams, respectively 
shown in Figure \ref{fig:sample_testset_tsne_t90_2d_3d}. 
The feature maps are output from the last convolutional layer of optimal model. 
Both 2D and 3D diagrams visualized from the initial count maps in test set are difficult to distinguish between GRBs and non-GRBs.
By contrast, the 2D and 3D diagrams obtained from feature maps shown distinct distribution and boundaries  
between GRBs and non-GRBs. 
This directly demonstrates the effectiveness of our deep learning method in classifying bursts.
For reckoning the hidden physical basis of classification, 
we embedded information of the SNR, duration, hardness ratio, and hardness ratio deviation of these GRBs into diagrams, shown in Figure \ref{fig:feature_testset_tsne_t90_2d_3d}, Figure \ref{fig:feature_testset_tsne_snr_2d_3d}, Figure \ref{fig:feature_testset_tsne_hardness_2d_3d}, and Figure \ref{fig:feature_testset_tsne_hardness_std_2d_3d}. 
We find such a black box that extracted 
features and classified decision are involved in these physical information. 
Supervised deep learning algorithms, with the input of the observed identification samples, can classify 
different categories into distinct clusters. 
The clusters show high correspondence with of GRBs and non-GRBs in high accuracy.
Visualizing analysis of count maps and feature maps by utilizing the t-SNE technique, 
a significant difference 
between the results of two classifications is observed. 
By inputting feature maps produced by ResNet-CBAM model, GRBs and non-GRBs 
are adequate clustered into different 
regions, which suggests the feature extraction of our trained optimal model is powerful and effective. 
For the input of GRBs that identified by Fermi, it is possible that some GRB samples may have been incorrectly classified as non-GRBs (FN samples).
By analyzing the classification results of the test set via the optimal model, we found that 
about 6\% samples of the GRB category were indeed misclassified. 
These embedded physical information could help us to understand the critical features 
of the identifying process of our optimal model. 
We find most of these FN samples were short in duration, low in significance, and abnormal in hardness deviation，which suggests that these observed traits have high weight ratio for decision
making of classification judgment. 
This suggests that these observed characters have a high weight ratio for decision making of classification judgment. 
\par
The optimal model is applied on one year observation data of Fermi/GBM, with 1963 candidates identified 
(see Figure \ref{fig:candi_category} and Table \ref{table:searched_event}). 
By comparing them with the trigger catalog of Fermi, we find most of the published GRBs were distinguished. 
A number of other types of triggers are also retrieved, showing that triggers are very complex and 
highly similar to GRB type within the existing data dimensions (count maps data). 
Therefore non-GRB type triggers may have been identified by our algorithm of binary classification. 
It suggests that upgrade of the data dimension in the future is necessary, such as adding of location 
information from multi-detectors observations, time-resolved spectral information, light curves in different 
energy channel, etc. 
None of the TGF events match unknown candidates, which probably due to their extremely short duration 
(typically 0.1\,ms \citep{fermi_gbm_fouth_catalog}) and their morphologies are completely 
different from GRBs. 
More than six hundred weak trigger signals, sub-threshold triggers, are not explicitly categorized by Fermi, 
they are too weak to distinguish, and rarely associate with our candidates.
Currently, our positive samples of training sets are no other than GRBs in Fermi burst catalog.
It is a disadvantage factor for identifying sub-threshold triggers. 
Suppose with the addition of such manual sub-threshold triggers to the training set, 
the trained model could provide stronger capability to extract features of weak 
triggers which would be the bursts from distant universe.
Admittedly sub-threshold GRBs are too intrinsically weak or viewed with unfavorable 
instrument geometry to initiate an onboard trigger of Fermi/GBM \citep{kocevski_2018_analysis_sub_grb}. 
Undeniably, it is worth identifying sub-threshold GRBs with a lot of efforts, of which are suggested as 
the counterparts of gravitational waves, fast radio bursts or other transient 
events \citep{sub_grb_for_GW2,sub_grb_for_GW,tohuvavohu_2020_bat_sub_grb_search}. 
The deep learning method could effectively uncover hidden and weak signals. 
In our candidates, most public sub-threshold GRBs of Fermi/GBM are precisely identified. 
Sufficient samples of sub-threshold GRBs would help us to train a more accurate algorithm 
with high efficiency than manual work \citep{zhang2018fast_CNN_detect_FRB,CNN_search_FRB_81}. 
In addition, the no matching candidates, totally are the 1558 unknown events, 
are potentially have the value of scientific research.
There are a number of additional bursts are    
discovered from SGR J1935+2154 beyond the Fermi/GBM trigger catalog 
by using the sub-threshold search algorithm, see Figure \ref{table:candi_sgr_match}.
There are 24 candidates from our optimal model are corresponding to these bursts. 
Through the SNR and localization analysis for the unknown candidates, we notice that
3 events apparently came from the SGR J1935+2154, and the spatial distribution of these 
localized candidates are uniform.
These offer unambiguous clue that our candidates would correlate with potential violent 
outbursts from astronomy objects. 
Subsequently deeper analysis would provide more associations between our candidates 
and other known bursts subsequently.
\par
The CNN provide higher sensitivity in many scenarios, but lack interpretability in prediction.
The visual analysis reveals that our model accurately identifies the 
burst characteristics of GRBs, which indicates that the model is accurate and effective. 
Compared to traditional trigger search algorithms implemented by peak detection, 
our approach makes more reasonable representation of the characteristics of GRBs. 
There are many types of burst phenomena besides GRB, such as fast radio bursts, 
soft gamma repeaters. 
Using multiple types of burst data in dataset to train the deep learning model, 
the visual analysis method enables a precise comparison and analysis of the 
characteristics and physical patterns of different bursts. 
The average elapsed time for the optimal model to classify each count map on 
GPU (NVIDIA GTX-1080Ti) is 8\,ms, which means that the model is able to identify 
GRBs in real time. 
In the future, our method allows to automate the quick look of scientific achievements 
with high precision, which is beneficial in guiding rapid follow-up observations. 
Our method could be adopted as a crucial procedure for the burst advocates (BA) to discover burst events, 
which assists researchers to improve the detection efficiency 
of GRBs or other burst events in further analysis. 
Transfer learning is a advance machine learning method that implements an 
existing model to a relevant task but only require tiny data set \citep{transfer_learning}. 
Gravitational wave high-energy Electromagnetic Counterpart All-sky Monitor (GECAM) satellites 
(i.e. GECAM-B and GECAM-C) were dedicated to monitoring Gamma-ray transients including 
GRBs and Soft Gamma Repeaters from the universe with considerable detection capability. 
With an orbital period of three years in orbit, 
the GECAM series have detected about two hundreds 
GRBs\footnote{\url{https://gcn.gsfc.nasa.gov/gcn/gecam_events.html}}, 
which were difficult to train large deep learning model for research.
Our deep learning models have great potential to be transferred into the data analysis pipelines of the GECAM telescope for distinguishing GRBs accurately. As more observations are available, more data will be collected to training our models continuously, which will be beneficial to effectively enhancing the generalization ability of models. Moreover, the feasibility of embedding tiny deep learning model into 
Satellite On-Board Computer (OBC) of Gamma-ray monitor telescope for onboard trigger distinguish  
is also worth exploring further. 
Take extremely early electromagnetic emission of GRBs as an example, e.g., 
the prompt optical emissions, the early optical flashes and the X-ray/ultraviolet flares, 
which are important probes for studying the physical origins and processes of 
GRBs \citep{follow_paper1, follow_paper2, follow_paper3, follow_paper4, follow_paper5}. 
These emissions are extremely brief and of high intensity, 
making challenges for observation and study. 
Real-time and accurate identification of trigger signals by deep learning models, 
and rapid transmission of information through satellite communication 
networks such as BAS of GECAM \citep{gecam_BAS} 
are crucial for follow-up observations by other telescopes in multi-wavelength bands. 
It enables researchers to capture multi-wavelength electromagnetic emissions as early as possible.

\section{Acknowledgements}
We would like to thank Dr. Xiang Ma, Dr. Riccardo Crupi, Dr. Shi-Jie Zheng, Dr. Zheng-De Zhang, 
and Dr. Shuo Xiao et al. for helpful discussion. 
This study is supported by the National Natural Science Foundation of 
China (Grant Nos. 12103055, U1938201, 12273042, 12133007, U1938108, 41827807, and 61271351), 
and by the National Key R\&D Program of China (2021YFA0718500).
This work is also partially supported by the Strategic Priority Research Program of the
CAS under grant No. XDA15360300. Z.-C. Z. is supported by the National Natural Science
Foundationof China (Grant No. 12233002). Bing li acknowledges support from the National
Astronomical Science Data Center Young Data Scientist Program (grant No. NADC2023YDS-04).

\section{Data Availability}
\label{sec:data_availability}
The code, data sets, and candidate list are available upon reasonable request.

\bibliographystyle{aasjournal}
\bibliography{paper.bib}

\appendix

\section{Model Evaluation Metrics}
\label{sec:metrics}
Generally, the performance of a deep learning model for classification is described by 
four metrics, namely, \textit{Accuracy}, \textit{Precision}, 
\textit{Recall}, and \textit{F1-score}, 
as follows
\begin{equation}
 Accuracy=\frac{TP+TN}{TP+TN+FP+FN}
\end{equation}
\begin{equation}
 Precision=\frac{TP}{TP+FP}
\end{equation}
\begin{equation}
 Recall=\frac{TP}{TP+FN}
\end{equation}
\begin{equation}
 F1\text{--}score=\frac{2\times Precision\times Recall}{Precision+Recall}
\end{equation}
where $ TP $, $ TN $, $ FP $ and $ FN $ represent the number of true positives,
true negatives, false positives and false negatives respectively. 

\section{Calculation of $T_{90}$ and SNR}
\label{sec:appendix_cal_snr_t90} 
We first fit the background using a polynomial function with order 1 for the 
summed light curve of triggered detectors. 
The background region is chosen from ($T_{begin}-15$\,s to $T_{begin}-5$\,s) 
and ($T_{end}+5$\,s to $T_{end}+15$\,s) which is relative to the burst period. 
The cumulative counts of each bin are plotted by using the background-subtracted light curve. 
The $T_{90}$ is calculated as $T_{90}=T_{95}-T_{5}$, where $T_{95}$ and $T_{5}$ are the 
times when 95 percent and 5 percent of the total counts of burst are obtained, respectively. 
Generally, we choose the range of $T_{90}$ as the source region of burst. 
We determine the SNR of the source region excess over background fluctuation, 
by using the following equations:
\begin{equation}
N_{bkg}=N^{'}_{bkg} \times \frac{T_{src}}{T_{bkg}}
\end{equation}
where $N^{'}_{bkg}$ and $T_{bkg}$ are the total counts and duration of the
background region, respectively. $T_{src}$ is the duration of the source region. 
Then the SNR can be estimated as:
\begin{equation}
SNR=\frac{ N_{net}  }{ \sigma_{N_{bkg}} }=\frac{ N_{src}-N_{bkg} }{ \sqrt{N_{bkg}} }
\end{equation}
where $N_{src}$ is the total counts in the source region.

\section{Localization algorithm of candidates}
\label{sec:appendix_locate} 
Based on the counts distribution of detectors in different directions, 
a Bayesian localization method is proposed to localize bursts using the  
Poisson data with Gaussian background (PGSTAT) profile likelihood \citep{YiZhao_LOC_Method,gecam_locate_bursts}. 
Due to the high-energy transients localization capability of this algorithm, 
we apply it to locate our candidates. 
The $T_{90}$ is used as the source interval of the candidates. 
We utilize the GBM Data Tools\footnote{GBM Data Tools: 
\url{https://fermi.gsfc.nasa.gov/ssc/data/analysis/rmfit/gbm_data_tools/gdt-docs/}.} 
to implement the polynomial fitting to the background. 
The estimated background of the source interval could be obtained by interpolation. 
Spectral data from the source interval of the 12 NaI detectors are used to perform fixed template localization. 
We adopt three fixed spectral templates (see also the spectral
templates of Table 4 in \citet{gecam_locate_bursts}) through the Bayesian localization method with PGSTAT profile likelihood. 
The maxima of the maximum PGSTAT profile likelihood is taken as the location result. 


\begin{deluxetable*}{cccc}[htbp]
\label{table:data set}
\tablecaption{
Description of data set.}
\tablehead{
Partition & Nu. of GRB & Nu. of non-GRB & Data Period Definition (UTC)}
\startdata
 Training set & 3082 & 6000 &  Jul 14, 2008 - Dec 31, 2014 \\ 
 Validation set & 1507 & 2000 & Jan 1, 2015 - Dec 31, 2017 \\
 Test set & 1741 & 2000 & Jan 1, 2018 - Jun 31, 2021 \\
\enddata
\end{deluxetable*}


\begin{deluxetable*}{ccccccc}[htbp]
\label{table:network metrics}
\tablecaption{The performance of the individual models on the training set, validation set and test set respectively.}
\tablehead{
Time Scale & Model & Dataset & \textit{Accuracy} (\%) & \textit{Precision} (\%) & \textit{Recall} (\%) & \textit{F1-score} (\%)}
\startdata
\multirow{9}*{256\,ms} & \multirow{3}*{\textit{plain}-CNN} & Training set & 99.37 & 99.90 & 98.24 & 99.06 \\ 
& & Validation set & 95.12 & 97.44 & 91.04 & 94.13 \\ 
& & Test set & 94.01 & 97.25 & 89.66 & 93.30 \\
\cline{2-7}
& \multirow{3}*{ResNet} & Training set & 98.95 & 99.93 & 96.98 & 98.43 \\ 
& & Validation set & 95.49 & 98.07 & 91.30 & 94.57 \\ 
& & Test set & 94.84 & 98.19 & 90.58 & 94.23 \\
\cline{2-7}
& \multirow{3}*{ResNet-CBAM} & Training set & 98.65 & 99.59 & 96.43 & 97.98 \\ 
& & Validation set & 95.83 & 97.88 & 92.30 & 95.01 \\ 
& & Test set & 95.08 & 98.08 & 91.21 & 94.52 \\
\midrule
\multirow{9}*{128\,ms} & \multirow{3}*{\textit{plain}-CNN} & Training set & 99.44 & 99.96 & 98.41 & 99.18 \\ 
& & Validation set & 95.83 & 98.22 & 91.97 & 94.99 \\ 
& & Test set & 95.26 & 98.99 & 90.75 & 94.69 \\
\cline{2-7}
& \multirow{3}*{ResNet} & Training set & 98.00 & 99.25 & 94.84 & 96.99 \\ 
& & Validation set & 96.17 & 98.51 & 92.50 & 95.41 \\ 
& & Test set & 95.29 & 98.69 & 91.09 & 94.74 \\
\cline{2-7}
& \multirow{3}*{ResNet-CBAM} & Training set & 98.73 & 99.83 & 96.43 & 98.10 \\ 
& & Validation set & 96.49 & 98.59 & 93.16 & 95.80 \\ 
& & Test set & 95.99 & \textbf{99.13} & 92.18 & 95.53 \\
\midrule
\multirow{9}*{64\,ms} & \multirow{3}*{\textit{plain}-CNN} & Training set & 99.18 & 99.60 & 97.98 & 98.78 \\ 
& & Validation set & 96.49 & 98.39 & 93.36 & 95.81 \\ 
& & Test set & 96.15 & 98.30 & 93.33 & 95.75 \\
\cline{2-7}
& \multirow{3}*{ResNet} & Training set & 99.21 & 99.53 & 98.15 & 98.84 \\ 
& & Validation set & 96.60 & 97.40 & 94.62 & 95.99 \\ 
& & Test set & 96.20 & 97.96 & 93.79 & 95.83 \\
\cline{2-7}
& \multirow{3}*{ResNet-CBAM} & Training set & 99.22 & 99.70 & 98.02 & 98.85 \\ 
& & Validation set & 96.94 & 98.20 & 94.62 & 96.30 \\ 
& & Test set & \textbf{96.57} & 98.61 & \textbf{93.96} & \textbf{96.23} \\
\enddata
\tablecomments{Bold text indicates the metrics for which the model performed best on the test set.}
\end{deluxetable*}

\begin{deluxetable*}{cccccc}[htbp]
\label{table:network_fusion}
\tablecaption{Summary of four metrics describing the performance of fused models.}
\tablehead{
Time Scale & Model & \textit{Accuracy} (\%) & \textit{Precision} (\%) & \textit{Recall} (\%) & \textit{F1-score} (\%)}
\startdata
\multirow{4}{*}{256\,ms} & \textit{plain}-CNN+ResNet              & 94.86   & 98.61   & 90.23   & 94.24 \\ 
& \textit{plain}-CNN+ResNet-CBAM        & 94.54   & 98.67   & 89.48   & 93.85 \\ 
& ResNet+ResNet-CBAM           & 95.40   & 98.51   & 91.49   & 94.87 \\ 
& \textit{plain}-CNN+ResNet+ResNet-CBAM & 95.10   & 97.44   & 91.90   & 94.59 \\
\midrule
\multirow{4}{*}{128\,ms} & \textit{plain}-CNN+ResNet              & 95.50   & \textbf{99.37}   & 90.92   & 94.96 \\ 
& \textit{plain}-CNN+ResNet-CBAM        & 95.53   & 99.24   & 91.09   & 94.99 \\
& ResNet+ResNet-CBAM           & 96.07   & 99.07   & 92.41   & 95.63 \\
& \textit{plain}-CNN+ResNet+ResNet-CBAM & 95.96   & 99.13   & 92.13   & 95.50 \\
\midrule
\multirow{4}{*}{64\,ms}  & \textit{plain}-CNN+ResNet              & 96.44   & 98.84   & 93.45   & 96.07 \\
& \textit{plain}-CNN+ResNet-CBAM        & 96.71   & 98.97   & 93.91   & 96.37 \\ 
& ResNet+ResNet-CBAM           & 96.60   & 98.67   & 93.96   & 96.26 \\
& \textit{plain}-CNN+ResNet+ResNet-CBAM & \textbf{96.81} & 98.85   & \textbf{94.25}   & \textbf{96.50} \\
\enddata
\tablecomments{Bolded text indicates that the model performs best on that metric.}
\end{deluxetable*}

\begin{deluxetable*}{cccc}[htbp]
\label{table:searched_event}
\tablecaption{Matching results of our searched events compared with Fermi/GBM published events. 
}
\tablehead{
Event & Our & Fermi & Searched Rate (\%)}
\startdata
GRB & 186 & 190 & 97.89 \\ 
SFLARE & 31 & 55 & 56.36 \\ 
SGR & 89 & 173 & 51.45 \\ 
LOCLPAR & 21 & 45 & 46.67 \\ 
UNCERT & 6 & 36 & 16.67 \\ 
TGF & 0 & 74 & 0 \\ 
\midrule
SubGRB & 8(9*) & 11 & 72.73(81.81*) \\ 
SubTrigger & 64(184*) & 642 & 9.96(28.66*) \\ 
\midrule
Unknown & 1558 & - & -  \\ 
\enddata
\tablecomments{The LOCLPAR, SFLARE, SGR, TGF and UNCERT events are the candidates corresponded
to the Fermi triggers, see \citep{fermi_catelog} for the specific definition of these triggers. 
The SubGRB and SubTrigger denote sub-threshold GRBs and sub-threshold triggers 
published by Fermi/GBM, respectively. 
The * indicates that only the GRB feature on a single detector matches the event. 
The unknown events indicate that these candidates are unable to match known events of Fermi.
}
\end{deluxetable*}

\begin{deluxetable*}{cccccc}[htbp]
\label{table:gbm_not_grb_list}
\tablecaption{
Comparison of the other unknown GRBs find by our model and various instrument.
}
\tablehead{GRB name & GCN number & Start of $T_{90}$ & $T_{90}$ & SNR & Detectors\\
& & (UTC) & (s) & ($\sigma$) & (n0-nb)
}
\startdata
GRB210708B & 30417(Swift/BAT) & 13:23:05.320288 & 29.57 & 10.3 & n3,n4,n7,n8 \\
GRB210724A & 30497(Swift/BAT) & 20:14:02.903808 & 24.26 & 15.4 & n9,na \\
GRB210930A & 30897(Swift/BAT) & 02:46:59.416032 & 7.17 & 10.7 & n0,n1,n2 \\
GRB220308A$^{*}$ & 31716(Swift/BAT), 31723(Konus-Wind), 31733(AGILE) & 05:43:09.182368 & 16.19 & 525.5 & n0-nb \\
GRB220403C$^{*}$ & 31824(Swift/BAT), 31907(Konus-Wind) & 10:13:53.125568 & 11.90 & 337.5 & n0-n5,n9,na,nb \\
\midrule
GRB210820A & 30664(Swift/BAT) & - & - & - & n9 \\
GRB210822A & 30677(Swift/BAT), 30678(GECAM), 30694(Konus-Wind) & - & - & - & n6 \\
GRB211022A & 30955(GECAM), 30968(AstroSat CZTI) & - & - & - & n8 \\
GRB211025A & 30986(Swift/BAT) & - & - & - & n3 \\
GRB211107B & 31057(Swift/BAT), 31086(Konus-Wind) & - & - & - & n6 \\
GRB211218A & 31261(MAXI/GSC) & - & - & - & n2 \\
GRB220117B & 31468(Swift/BAT) & - & - & - & n8 \\
GRB220219B & 31646(Konus-Wind) & - & - & - & n2 \\
GRB220302A & 31661(Swift/BAT) & - & - & - & n1 \\
GRB220310A & 31725(MAXI/GSC), 31748(Konus-Wind) & - & - & - & na \\
GRB220404A & 31828(Swift/BAT) & - & - & - & n1 \\
GRB220412A & 31881(Swift/BAT) & - & - & - & n8 \\
GRB220506A & 32057(Konus-Wind) & - & - & - & n3 \\
GRB220514B & 32046(MAXI/GSC), 32056(Konus-Wind) & - & - & - & n6 \\
GRB220519A & 32073(Konus-Wind) & - & - & - & n2 \\
GRB220623A & 32243(Swift/BAT), 32246(AstroSat CZTI), 32258(Konus-Wind) & - & - & - & n1 \\
\enddata
\tablecomments{
Matching results of our searched events compared with GRBweb and GCN circular published GRBs which 
not listed in Fermi/GBM catalog. \\
\footnotesize{$^{*}$The trigger time of the two GRBs, GRB220308A and GRB220403C, are 
close with GRB220308233 and GRB220403424, respectively. Perhaps they are the same burst probably.}\\
}

\end{deluxetable*}

\begin{deluxetable*}{cccccc}[htbp]
\label{table:candi_sgr_match}
\tablecaption{
The bursts of SGR J1935+2154 in various researches and the comparison 
result.
}
\tablehead{Time & Lin$^{a}$ & Lin$^{b}$ & Zou$^{c}$ & Ibrahim$^{d}$ & Xie$^{e}$ }
\startdata
07/05/2014-08/27/2016 & 112 & - & 72 & - & - \\ 
10/04/2019-05/21/2020 & - & 148 & 152 & - & - \\ 
05/21/2020-07/01/2021 & - & - & 23 & - & - \\ 
\midrule
07/01/2021-09/09/2021 & - & - & 11 & - & - \\ 
09/09/2021-10/01/2021 & - & - & 90 & 79 & - \\ 
10/01/2021-10/12/2021 & - & - & 5 & - & - \\ 
11/07/2021-01/18/2022 & - & - & - & - & 145 \\ 
01/18/2022-06/30/2022 & - & - & - & - & - \\ 
\midrule
matched with Fermi trigger & 62 & 44 & 167 & 67 & 67 \\ 
matched with Fermi subTrigger & - & - & 8 & 2 & 10 \\ 
\midrule
matched with our unknown events & - & - & 15 & 7 & 12 \\ 
\enddata
\tablecomments{
The time periods and bursts number in such different periods are 
shown in columns of the first two part. 
The corresponding literatures are 
\footnotesize{$^a$ \citet{sgr_list_linlin}}, 
\footnotesize{$^b$ \citet{sgr_list_linlin2}}, 
\footnotesize{$^c$ \citet{sgr_list_zoujinhang}},
\footnotesize{$^d$ \citet{sgr_list_Rehan}},
\footnotesize{$^e$ \citet{sgr_list_xieshenglun}}. 
A lot of bursts are overlapped in same period,
such as Lin et al. 2020a and Zou et al. 2021. 
The third part represents the matched number of Fermi triggers and subTriggers with those articles in 5 second interval. 
The last row shows the number of those published bursts that confirms our unknown events. 
Of the 34 matched bursts, 10 events are overlapped. 
}
\end{deluxetable*}

\begin{deluxetable*}{ccccccccccc}[htbp]
\label{table:candi_sgr_match_list}
\tablecaption{
The comparison result of the bursts found by our model and 
referred researches from SGR J1935+2154. 
}
\tablehead{ID & Zou$^{a}$ & Ibrahim$^{b}$ & Xie$^{c}$ & Start of $T_{90}$ & $T_{90}$ & SNR & RA & Dec & err & Detectors\\
 & & & & (UTC) & (s) & ($\sigma$) & (deg) & (deg) & (deg) & (n0-nb)}
\startdata
210805A & Y & N & N & 00:08:55.888314 & 0.19 & 162.7 & 296.2 & 24.3 & 0.8 & n3,n4,n6,n7,n8,nb \\
210910A & Y & N & N & 01:04:32.065138 & 2.37 & 9.6 & 297.8 & 22.9 & 9.2 & n6,n7 \\
210910B & Y & N & N & 01:08:40.337138 & 2.75 & 18.5 & 294.4 & 27.0 & 4.3 & n6,n7,n9,na \\
210910C & Y & Y & N & 01:17:18.993138 & 0.19 & 86.8 & 295.7 & 24.9 & 1.0 & n0,n1,n6,n9,na,nb \\
210910D & Y & Y & N & 01:18:53.505138 & 1.15 & 19.9 & 295.7 & 23.9 & 2.9 & n1,n6,n9,na \\
210910E & Y & N & N & 01:21:48.321138 & 1.28 & 16.9 & 297.0 & 19.6 & 2.7 & n0,n1,n6,n9,na,nb \\
210910F & Y & N & N & 01:31:39.921138 & 0.45 & 30.3 & 296.3 & 16.5 & 2.6 & n9,na,nb \\
210910G & Y & N & N & 01:34:18.737138 & 0.26 & 48.6 & 299.1 & 19.0 & 1.9 & n9,na,nb \\
210910H & Y & Y & N & 02:44:34.017352 & 0.19 & 93.6 & 295.8 & 26.4 & 1.4 & n4,n6,n7,n9,na,nb \\
210911A & Y & N & N & 17:10:48.320750 & 0.90 & 28.4 & 300.3 & 14.5 & 3.0 & n9,na,nb \\
210911B & Y & Y & N & 20:13:40.400030 & 0.26 & 75.1 & 296.2 & 13.8 & 2.1 & n1,n9,na \\
210911C & Y & Y & N & 15:06:43.169002 & 0.38 & 476.4 & 293.9 & 23.1 & 0.6 & n1-nb \\
210911D & Y & Y & N & 15:17:45.217002 & 0.38 & 305.7 & 300.3 & 16.9 & 0.6 & n0-n8,na,nb \\
210911E & Y & Y & N & 15:32:33.345002 & 0.32 & 74.5 & 297.9 & 15.8 & 1.3 & n0,n1,n6,n9,na,nb \\
210911F & Y & N & N & 15:34:39.793002 & 3.71 & 16.6 & 302.6 & 21.0 & 5.1 & n9,na \\
220112A & N & N & Y & 01:08:01.457448 & 0.26 & 102.7 & 295.3 & 19.6 & 1.7 & n0,n1,n3,n4,n5 \\
220112B & N & N & Y & 18:12:02.256072 & 2.75 & 9.0 & 287.0 & 17.8 & 10.8 & n0,n1,n2,n5 \\
220113A & N & N & Y & 08:34:19.666714 & 0.58 & 6.2 & 305.9 & 27.7 & 2.7 & n1,n7 \\
220113B & N & N & Y & 15:13:21.088072 & 5.06 & 8.3 & 293.8 & 14.3 & 10.6 & n0,n3 \\
220114A & N & N & Y & 20:46:32.873062 & 0.45 & 73.0 & 301.3 & 27.5 & 2.3 & n1,n2,n5 \\
220115A & N & N & Y & 08:25:55.152818 & 1.54 & 71.1 & 293.5 & 18.0 & 1.5 & n0-n7,na \\
220115B & N & N & Y & 08:45:40.592818 & 0.13 & 105.8 & 299.2 & 21.5 & 1.6 & n1,n3,n4,n5 \\
220115C & N & N & Y & 13:13:05.280634 & 3.39 & 15.3 & 290.3 & 21.4 & 5.3 & n0,n3 \\
220115D & N & N & Y & 19:19:45.936170 & 0.19 & 20.3 & 297.4 & 25.1 & 5.3 & n0,n1 \\
\enddata
\tablecomments{
The first column is the id we compiled for the SGR event.
The corresponding literatures are 
\footnotesize{$^a$ \citet{sgr_list_zoujinhang}},
\footnotesize{$^b$ \citet{sgr_list_Rehan}},
\footnotesize{$^c$ \citet{sgr_list_xieshenglun}}. 
The calculations of $T_{90}$ and SNR are shown in Appendix \ref{sec:appendix_cal_snr_t90}. 
The localization algorithm is presented in Appendis \ref{sec:appendix_locate}. 
We use the localization error within 1 $\sigma$ as the equivalent radius of the position region, see column 'err'. 
The last column represents the detectors where our model find that the burst feature matched the trigger time of the SGR. 
}

\end{deluxetable*}


\begin{longrotatetable}
\begin{deluxetable}{lcccccccclcccccccc}
\centerwidetable 
\tabletypesize{\footnotesize}
\tabcolsep=0.1cm
\label{table:candi_snr_greater_5}
\tablecaption{Candidates with SNR $ \ge 5\,\sigma$ of unknown events. 
}
\tablehead{Can.ID & Start of $T_{90, F}$ & $T_{90, F}$ & Start of $T_{90}$ & $T_{90}$ & SNR & RA & Dec & err & Can.ID & Start of $T_{90, F}$ & $T_{90, F}$ & Start of $T_{90}$ & $T_{90}$ & SNR & RA & Dec & err\\ 
& (UTC) & (s) & (UTC) & (s) & ($\sigma$) & (deg) & (deg) & (deg) & & (UTC) & (s) & (UTC) & (s) & ($\sigma$) & (deg) & (deg) & (deg)}
\startdata
\textbf{210702A}$^{ }$ & 00:23:44.208415 & 20.03 & 00:23:51.120406 & 11.90 & 17.7 & 221.2 & 19.6 & 5.6 & \textbf{220306B}$^{ }$ & 07:57:16.097252 & 25.66 & 07:57:30.753246 & 11.46 & 10.5 & 108.1 & 64.4 & 9.8 \\ 
\textbf{210703F}$^{ }$ & 06:29:43.865561 & 58.43 & 06:29:50.777554 & 44.93 & 9.0 & 16.1 & 9.9 & 8.5 & \textbf{220309B}$^{ }$ & 10:23:18.833991 & 15.42 & 10:23:20.497976 & 16.58 & 6.7 & 342.5 & 10.3 & 5.0 \\ 
\textbf{210706C}$^{\circledast}$ & 21:48:30.000608 & 42.50 & 21:48:39.792608 & 49.47 & 74.1 & 105.4 & 25.4 & 0.6 & \textbf{220309E}$^{ }$ & 22:06:20.736690 & 34.50 & 22:06:33.536680 & 16.96 & 5.6 & 282.9 & 60.1 & 21.1 \\ 
\textbf{210710D}$^{\dagger}$ & 23:41:57.677494 & 16.26 & 23:42:07.725490 & 0.19 & 36.0 & 295.5 & 21.0 & 3.3 & \textbf{220310A}$^{\star}$ & 03:00:56.816929 & 17.34 & 03:00:42.480918 & 30.85 & 46.6 & 136.2 & 9.1 & 1.9 \\ 
\textbf{210718A}$^{\dagger\bullet}$ & 02:16:39.776219 & 32.77 & 02:16:57.056216 & 0.19 & 14.3 & 14.3 & 6.7 & 6.7 & \textbf{220311F}$^{\bullet}$ & 18:29:56.893393 & 26.11 & 18:30:01.373382 & 22.40 & 5.5 & 290.0 & 22.5 & 29.6 \\ 
\textbf{210720G}$^{\dagger}$ & 13:19:10.001086 & 25.73 & 13:19:28.241086 & 0.19 & 13.7 & 234.3 & -26.8 & 8.6 & \textbf{220313B}$^{\circledast}$ & 04:13:32.545743 & 20.35 & 04:13:36.321734 & 16.70 & 41.3 & 354.9 & -1.1 & 2.1 \\ 
\textbf{210728B}$^{\dagger\bullet}$ & 01:43:18.432015 & 16.00 & 01:43:28.736012 & 0.32 & 9.0 & 30.5 & 24.9 & 6.8 & \textbf{220314A}$^{\bullet}$ & 01:09:18.048336 & 41.66 & 01:09:22.912336 & 40.38 & 36.7 & 92.7 & 65.9 & 0.6 \\ 
\textbf{210807D}$^{ }$ & 23:37:41.392846 & 61.95 & 23:37:44.592846 & 30.46 & 8.6 & 61.2 & 81.6 & 7.5 & \textbf{220314C}$^{\bullet}$ & 02:46:49.057621 & 41.09 & 02:47:10.881610 & 17.86 & 6.3 & 150.4 & -13.3 & 24.8 \\
\enddata
\tablecomments{
We use the localization error within 1 $\sigma$ as the equivalent radius of the position region, see column 'err'. 
\\
\footnotesize{$^\dagger$The event is considered to be a likely spark event after manual inspection.}\\
\footnotesize{$^\star$The event is part of a long burst.}\\
\footnotesize{$^\bullet$The event is located in the earth occlusion region.}\\
\footnotesize{$^\circledast$According to the location, the signal of the event come from the sun.}\\}
\end{deluxetable}
\end{longrotatetable}

\begin{longrotatetable}
\begin{deluxetable}{lcccccccclcccccccc}
\centerwidetable
\tabletypesize{\footnotesize}
\tabcolsep=0.1cm
\label{table:candi_snr_less_5}
\tablecaption{Candidates with SNR $ \textless 5\,\sigma$ of unknown events.}
\tablehead{Can.ID & Start of $T_{90, F}$ & $T_{90, F}$ & Start of $T_{90}$ & $T_{90}$ & SNR & RA & Dec & err & Can.ID & Start of $T_{90, F}$ & $T_{90, F}$ & Start of $T_{90}$ & $T_{90}$ & SNR & RA & Dec & err\\ 
& (UTC) & (s) & (UTC) & (s) & ($\sigma$) & (deg) & (deg) & (deg) & & (UTC) & (s) & (UTC) & (s) & ($\sigma$) & (deg) & (deg) & (deg)}
\startdata
\textbf{210701A}$^{ }$ & 19:34:57.841309 & 24.06 & 19:35:10.129298 & 9.79 & 3.9 & 96.1 & -41.6 & 11.2 & \textbf{211204A}$^{\dagger}$ & 03:37:21.798223 & 17.22 & 03:37:33.766212 & 0.13 & 2.4 & 295.0 & 46.4 & 11.0 \\ 
\textbf{210703B}$^{ }$ & 01:21:26.414169 & 21.44 & 01:21:35.182160 & 7.68 & 4.7 & 359.4 & 2.0 & 14.1 & \textbf{211208D}$^{ }$ & 16:32:38.046245 & 14.66 & 16:32:45.918242 & 4.86 & 3.7 & 324.2 & -44.6 & 27.0 \\ 
\textbf{210711A}$^{ }$ & 03:08:21.840390 & 28.80 & 03:08:30.736388 & 9.79 & 2.9 & 13.2 & 12.9 & 25.1 & \textbf{211213D}$^{\bullet}$ & 18:39:03.247260 & 29.50 & 18:39:19.951260 & 6.85 & 4.3 & 234.6 & 31.3 & 50.1 \\ 
\textbf{210713D}$^{\bullet}$ & 19:22:53.313285 & 36.48 & 19:23:16.353282 & 2.94 & 4.7 & 81.5 & -40.1 & 24.1 & \textbf{211216D}$^{\bullet}$ & 16:59:06.048041 & 42.30 & 16:59:08.800030 & 25.66 & 4.6 & 258.8 & -6.2 & 14.7 \\ 
\textbf{210717E}$^{ }$ & 23:46:54.655941 & 27.97 & 23:47:04.831940 & 4.80 & 3.1 & 351.8 & 36.8 & 53.4 & \textbf{211222C}$^{ }$ & 12:14:10.848777 & 25.28 & 12:14:08.416772 & 24.77 & 4.3 & 153.8 & -13.9 & 23.7 \\ 
\textbf{210720A}$^{\bullet}$ & 02:58:04.383355 & 19.33 & 02:58:13.983344 & 7.17 & 4.2 & 208.7 & 27.9 & 13.2 & \textbf{211222E}$^{\bullet}$ & 14:27:48.160643 & 46.66 & 14:27:56.416638 & 3.26 & 3.7 & 89.2 & 36.1 & 61.9 \\ 
\textbf{210720B}$^{ }$ & 03:12:46.833447 & 27.97 & 03:12:52.017438 & 10.94 & 4.4 & 126.5 & -66.2 & 45.9 & \textbf{211226C}$^{ }$ & 08:21:01.568572 & 16.38 & 08:21:11.168562 & 2.43 & 4.6 & 4.7 & 7.7 & 38.0 \\ 
\textbf{210720I}$^{ }$ & 18:01:02.032114 & 18.24 & 18:01:08.496112 & 5.44 & 2.4 & 267.9 & -55.5 & 26.7 & \textbf{211226D}$^{\bullet}$ & 08:22:08.048565 & 27.14 & 08:22:16.432562 & 4.74 & 2.8 & 251.0 & 63.7 & 43.8 \\
\enddata
\tablecomments{
We use the localization error as the equivalent radius of the error region, see column 'err'. 
\\
\footnotesize{$^\dagger$The event is considered to be a likely spark event after manual inspection.}\\
\footnotesize{$^\star$The event is part of a long burst.}\\
\footnotesize{$^\bullet$The event is located in the earth occlusion region.}\\
\footnotesize{$^\circledast$According to the location, the signal of the event come from the sun.}\\}
\end{deluxetable}
\end{longrotatetable}

\begin{longrotatetable}
\begin{deluxetable}{lcclcclcclcc}
\centerwidetable
\label{table:candi_snr_no}
\tablecaption{Candidates without significant burst signal of unknown events.}
\tablehead{Can.ID & Start of $T_{90, F}$ & $T_{90, F}$ & Can.ID & Start of $T_{90, F}$ & $T_{90, F}$ & Can.ID & Start of $T_{90, F}$ & $T_{90, F}$& Can.ID & Start of $T_{90, F}$ & $T_{90, F}$\\
& (UTC) & (s) & & (UTC) & (s)& & (UTC) & (s)}
\startdata
\textbf{210702B} & 13:40:13.633995 & 35.90 & \textbf{211005E} & 07:06:08.320430 & 21.44 & \textbf{220112B} & 12:23:07.248419 & 24.58 & \textbf{220414B} & 03:44:13.700404 & 21.12 \\ 
\textbf{210703A} & 00:22:14.944398 & 19.26 & \textbf{211005F} & 07:14:45.808425 & 53.50 & \textbf{220112C} & 13:15:45.392294 & 34.69 & \textbf{220414C} & 04:13:16.673906 & 34.82 \\ 
\textbf{210703C} & 02:53:57.376001 & 28.54 & \textbf{211005G} & 08:34:32.497621 & 57.15 & \textbf{220112D} & 13:20:54.208282 & 37.70 & \textbf{220414D} & 12:46:42.080356 & 29.82 \\ 
\textbf{210703D} & 03:14:50.528868 & 29.57 & \textbf{211006A} & 04:24:54.292826 & 32.77 & \textbf{220112E} & 13:21:46.816282 & 13.12 & \textbf{220415A} & 01:57:08.706212 & 39.23 \\ 
\textbf{210703E} & 04:48:31.556519 & 22.27 & \textbf{211006B} & 06:10:06.523777 & 28.10 & \textbf{220112G} & 15:52:03.569711 & 30.40 & \textbf{220415B} & 04:39:53.377163 & 18.37 \\ 
\textbf{210703G} & 06:38:56.937562 & 25.34 & \textbf{211006C} & 08:07:20.416088 & 25.86 & \textbf{220113A} & 00:45:24.454114 & 30.72 & \textbf{220415C} & 04:43:26.849162 & 23.81 \\ 
\textbf{210703H} & 08:36:49.824151 & 19.65 & \textbf{211006D} & 09:45:22.871300 & 37.12 & \textbf{220113B} & 03:39:21.612292 & 31.42 & \textbf{220415D} & 07:39:42.691070 & 15.81 \\ 
\textbf{210703I} & 13:31:50.032010 & 22.21 & \textbf{211006E} & 12:00:50.000454 & 36.42 & \textbf{220113C} & 03:41:29.212295 & 16.70 & \textbf{220415E} & 10:09:48.448022 & 15.23 \\
\enddata
\tablecomments{
(This table is available in its entirety in machine-readable form.)}
\end{deluxetable}
\end{longrotatetable}

\end{CJK*}
\end{document}